\documentclass{natureprintstyle}

\usepackage{graphicx}
\usepackage{journals}
\usepackage[10.1038/s41550-021-01470-5]{natast_emu}

\title{Transforming gas-rich low-mass disky galaxies into ultra-diffuse galaxies by ram pressure}

\author{Kirill~A.~Grishin$^{1,2,8}$, Igor~V.~Chilingarian$^{3,1,8}$, 
Anton~V.~Afanasiev$^{4,5,1,8}$, Daniel~Fabricant$^{3}$, 
Ivan~Yu.~Katkov$^{6,1}$, Sean~Moran$^{3}$, Masafumi~Yagi$^{7}$}

\begin{document}

\twocolumn[
\begin{@twocolumnfalse}
\maketitle

\begin{abstract}
Faint extended elliptically-shaped ultra-diffuse galaxies and slightly brighter and more compact dwarf elliptical and lenticular stellar systems are common in galaxy clusters. Their poorly constrained evolutionary paths can be studied by identifying young ultra-diffuse galaxy and dwarf elliptical analogs populated with bright, massive stars. Using data mining we identified 11 such low-mass ($2\times10^8<M_*<2\times10^9 M_{\odot}$) galaxies with large half-light radii ($2.0<r_e<5$~kpc) and recently quenched star formation in the Coma and Abell~2147 galaxy clusters. All galaxies happen to have ram-pressure-stripped tails with signs of current or recent star formation. Deep spectroscopic observations revealed rotating stellar discs containing 70--95\% dark matter by mass. A large fraction of the disc stars (10--60\%) formed in intense star bursts 180--970~Myr ago, probably triggered by ram pressure. Observed global gradients of stellar age corroborate this scenario. Passive evolution in the next 10~Gyr will transform 9 of the 11 galaxies into ultra-diffuse galaxies. If we assume a constant rate of galaxy infall, 44$\pm$16~\%\ of the most luminous present-day ultra-diffuse galaxies in Coma must have formed via ram pressure stripping of disky progenitors.
\end{abstract}

\end{@twocolumnfalse}
]

\thispagestyle{firstpagestyle}

\mathfootnote{
\begin{affiliations}
 \item Sternberg Astronomical Institute, M.V. Lomonosov Moscow State University, 13 Universitetsky prospect, Moscow, 119992, Russia
 \item Department of Physics, M.V. Lomonosov Moscow State University, 1 Vorobyovy Gory, Moscow, 119991, Russia
 \item Center for Astrophysics -- Harvard and Smithsonian, 60 Garden St. MS09, Cambridge MA 02138, USA
 \item Universit\'{e} de Paris, CNRS, Astroparticule et Cosmologie, F-75013 Paris, France 
 \item LERMA, Observatoire de Paris, PSL Research University, CNRS, F-75014 Paris, France
 \item New York University Abu Dhabi, 129188, Saadiyat Island, Abu Dhabi, United Arab Emirates
 \item Subaru Telescope, National Astronomical Observatory of Japan, 2-21-1 Osawa, Mitaka, Tokyo 181-8588, Japan
 \item[*] These authors contributed equally to this work
\end{affiliations}
}

\vskip -4.8mm
\noindent
\lettrine[lines=3, lhang=0.15]{T}{\:}he first systematic studies of the two nearest galaxy clusters (Virgo and Fornax) three decades ago\cite{SB84,FS88} revealed large populations of low-luminosity ($L_B \lesssim 3\times10^9 L_{\odot}$) early-type galaxies, including some very extended examples ($r_e>5~kpc$). Recently, dozens\cite{2015ApJ...798L..45V} and then hundreds\cite{Koda15} of similar extended ``ultra-diffuse'' galaxies (UDGs) were found in the more distant ($d=99$~Mpc)\cite{Saulder+16} and populous Coma cluster and also in galaxy groups\cite{2017MNRAS.468.4039R}. These UDGs include several extreme examples with the size but $<$1\%\ of the stellar mass of the Milky Way. Initially some UDGs were thought to be ``failed galaxies''\cite{2015MNRAS.452..937Y}: massive dark matter halos that formed few stars following rapid gas loss at an early evolutionary stage. This gas loss could arise from the cluster environment, including ram pressure stripping by hot intracluster gas\cite{1972ApJ...176....1G} and tidal interactions\cite{1996Natur.379..613M} or from internal processes such as supernovae (SN) driven winds\cite{1986ApJ...303...39D}.  Gas loss from a low-mass star-forming galaxy will quench star formation, allowing the galaxy to evolve into a quiescent early-type system. Recent observations\cite{2018MNRAS.478.2034R} and simulations\cite{2019MNRAS.485..382C,2019MNRAS.487.5272J} suggest that a dwarf galaxy can be puffed up by SN feedback or tidal interactions and transformed into a UDG. It is unclear whether UDGs and dEs belong to the same family\cite{2018RNAAS...2a..43C,2019ApJ...884...79C} and share similar evolutionary paths or if they represent different galaxy types with different formation scenarios. In-depth observational studies of UDGs are challenging because their surface brightnesses are significantly below that of the night sky. This fundamental limitation can be overcome by identifying recently quenched low-mass galaxies, still containing young stars, that will passively evolve into dEs or UDGs. 

\begin{figure*}[h!]
	\centering
	\includegraphics[width=\hsize]{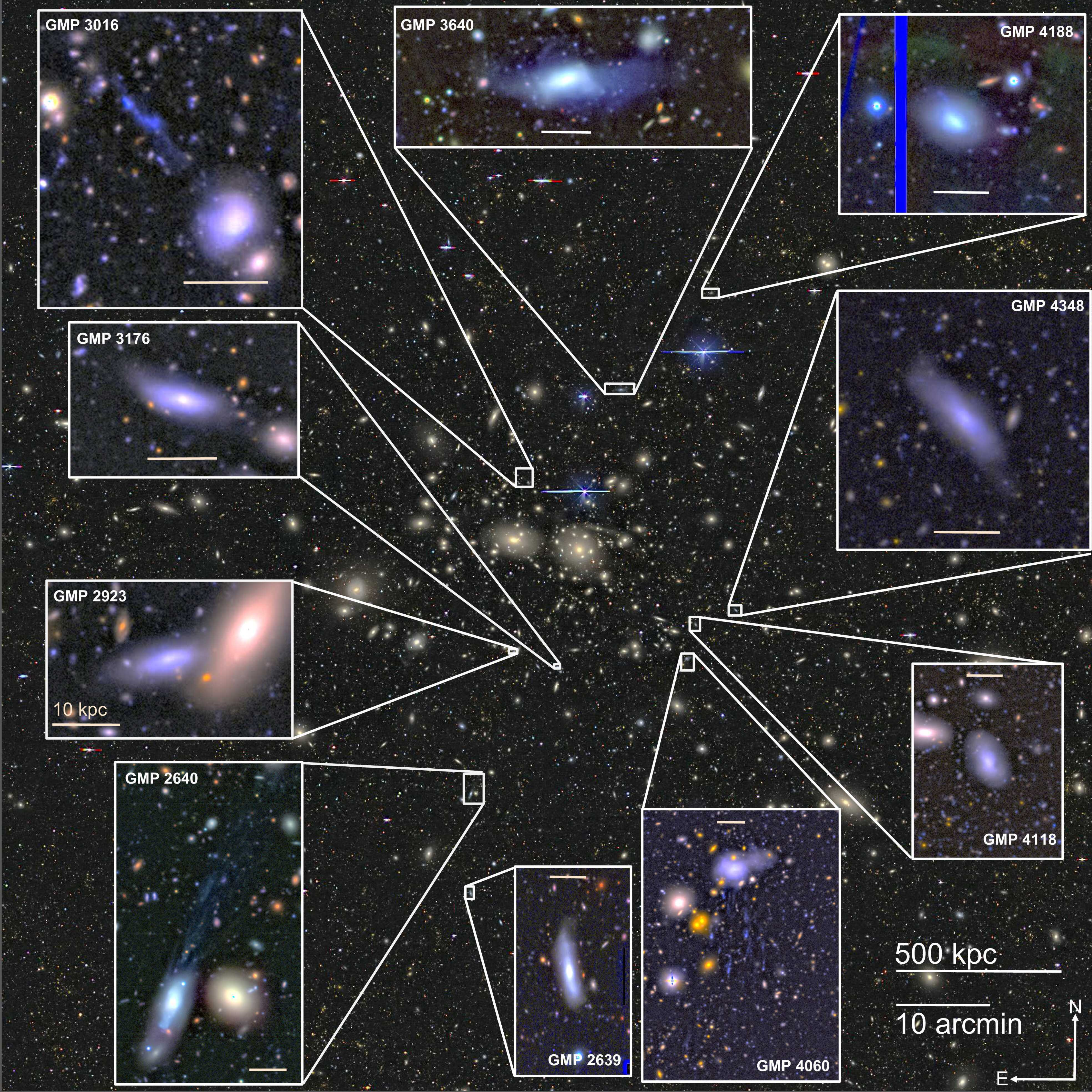}
		\caption{{\fontseries{b}\selectfont Positions of Coma cluster members selected from RCSED and one additional galaxy, GMP~3016.\fontseries{m}\selectfont} The insets showing zoomed-in galaxy images were constructed from Subaru and Canada-France-Hawaii Telescope (CFHT) optical data in the $u*$, $B$, $V$ and $R$ bands. A scale bar corresponding to 10~kpc is shown inside each inset. 
		In GMP~2640 the image reveals a blue clump of recently formed stars south-east of the centre which was masked for dynamical modelling purposes. 
		}
		\label{psg_coma}
\end{figure*}

\begin{figure*}[h!]
	\centering
	\includegraphics[width=0.7\hsize]{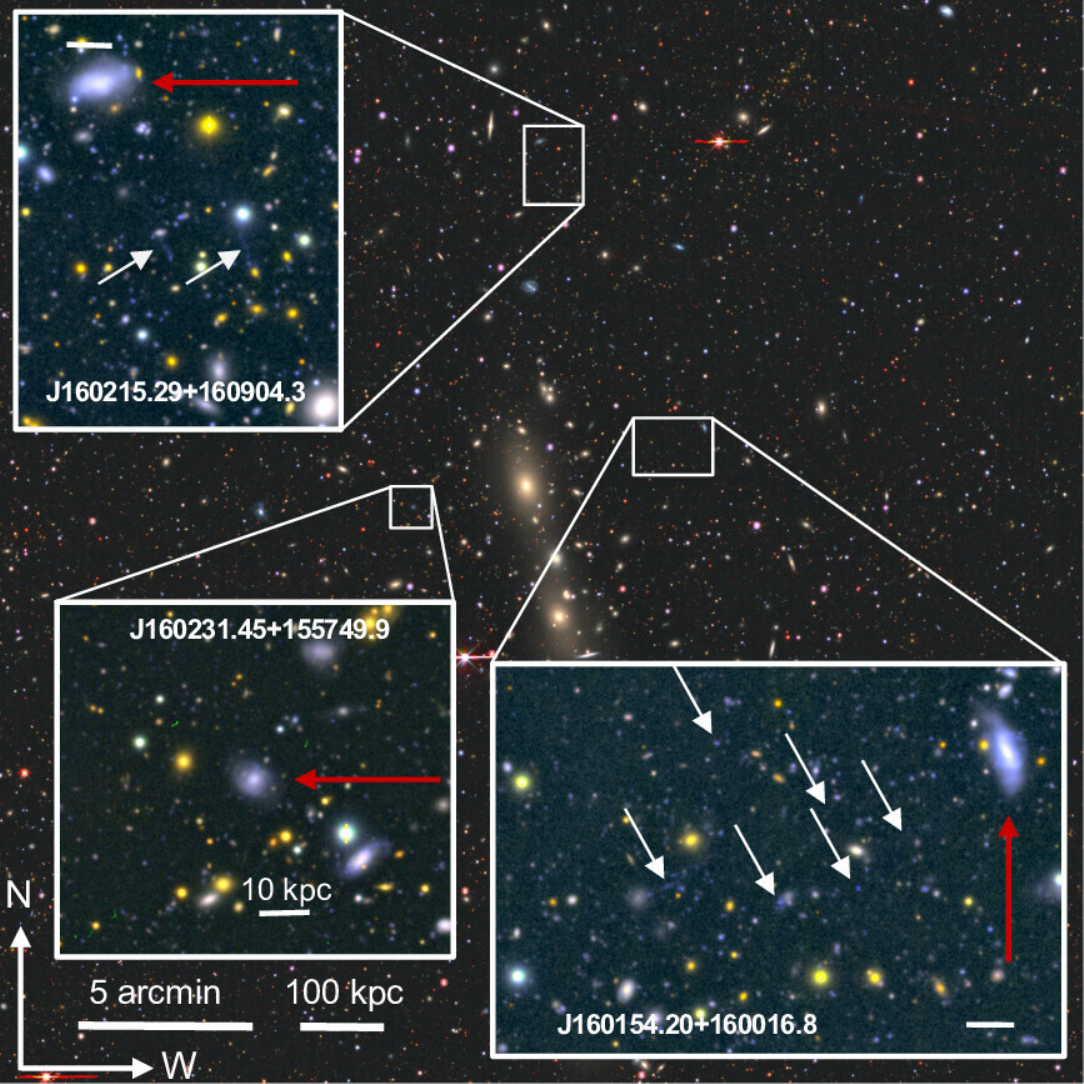}
		\caption{{\fontseries{b}\selectfont Positions of the two Abell~2147 cluster members selected from RCSED and an additional photometrically selected PSG, J160231.45+155749.9.\fontseries{m}\selectfont} The insets showing zoomed-in images for individual galaxies (marked by red arrows) are constructed from $g$, $r$, and $i$ broadband HSC images. The two galaxies from the main sample exhibit tails of stripped material. We mark the positions of clumps in the tails with white arrows.}
		\label{psg_a2147}
\end{figure*}

Data mining the multi-wavelength Reference Catalog of Spectral Energy Distribution of galaxies (RCSED\cite{RCSED}) revealed 12 blue but non-starforming galaxies (mean stellar age $<1.5$~billion years) with half-light radii $r_{e}$ between 2.0~kpc and 5.2~kpc.  The RCSED galaxies are part of the main galaxy sample from the 7th Data Release of the Sloan Digital Sky Survey\cite{SDSS_DR7},  and form a nearly complete magnitude-limited sample within the SDSS  footprint. Nine galaxies reside in the Coma cluster (Figure~\ref{psg_coma}), two are Abell~2147 cluster ($d=156$~Mpc) members (Figure~\ref{psg_a2147}), and one is a group member. Four of the nine Coma cluster members were previously classified as ``blue k+a'' post-starburst galaxies (PSGs)\cite{2004ApJ...601..197P} based on their integrated-light spectra. However, a recent starburst is needed to properly reproduce the spectra and colours of the remaining galaxies, so they are also PSGs. Two Coma galaxies (GMP~4060 and GMP~2923) are known\cite{2010AJ....140.1814Y} to exhibit spectacular tails of material stripped by the ram pressure of the hot intracluster medium\cite{1972ApJ...176....1G} with  H$\alpha$ emission suggesting current star formation.  Their discs, unlike most other galaxies in the the Subaru SuprimeCam H$\alpha$ survey of the Coma cluster, are not starforming. Young stars are UV-bright and expose a 250~kpc long tail in the most massive galaxy in our sample, GMP~2640\cite{2010MNRAS.408.1417S}. Visual inspection of Subaru HyperSupremeCam images revealed filamentary tails in both Abell 2147 members (Figure~\ref{psg_a2147}). Deep broad-band and H$\alpha$ images of the remaining Coma galaxies also expose faint low-contrast structures near them (Extended Data Figure~\ref{psg_suspected_tails}) that are likely the leftovers of ram-pressure-stripped tails significantly dimmed because of stellar evolution.

None of the tails point away from the cluster centre suggesting that these galaxies have orbits with significantly tangential shapes. The galaxies with the brightest tails have the youngest mean stellar ages of the 11 PSGs. The lack of star formation in their discs suggests that they could be classified as ``post-jellyfish'' galaxies. Similar, but typically more massive ``jellyfish'' systems\cite{2017ApJ...844...48P} have active star formation in their discs and tails\cite{2017ApJ...846...27G}. Two similar low-mass quenched PSGs with impressive tails, IC~3418\cite{2014ApJ...780..119K} and VCC~1249\cite{2012A&A...543A.112A} in the nearby Virgo cluster are believed\cite{2014ApJ...780..119K} to have been recently transformed by ram pressure stripping from star forming galaxies into quiescent dwarfs. However, at the Coma cluster distance both galaxies would have fallen below the SDSS magnitude limit. Prior to stripping, ram pressure can compress gas and induce a burst of star formation\cite{1999ApJ...516..619F,2009A&A...499...87K}, consistent with the PSG classification of our galaxies.

\section*{Results}

We observed the 11 low-mass PSGs in the Coma and Abell~2147 clusters using the high-throughput multi-object Binospec spectrograph\cite{2019PASP..131g5004F} operated at the 6.5-m MMT. GMP~3640 was also observed with the 8~m Gemini-North telescope using the GMOS-N spectrograph. We added five additional galaxies to the sample. These include GMP~3016, the remaining known Coma cluster member with H$\alpha$ detected in the tail but not in the main disc\cite{2010AJ....140.1814Y}. GMP~3016 is 0.03~mag below the SDSS magnitude limit and would have been excluded by the original selection criteria. We also included four additional galaxies without SDSS spectra, three fainter UDGs in the Coma cluster with mean stellar ages $<2$~Gyr\cite{2019ApJ...884...79C} and one photometrically selected faint PSG in Abell~2147. 

To assess the current properties of PSGs and predict their evolution, we adopted a simple physically motivated scenario for their star formation histories (SFHs). This scenario includes a period of constant star formation rate (SFR) that started 12~Gyr ago and a more recent ram pressure-induced starburst followed by the shutoff of star formation\cite{PSGM_IAU355}.  

\begin{figure*}[h!]
\centering
\includegraphics[trim=0.0cm 2.1cm 0.0cm 0.8cm, width=0.82\hsize]{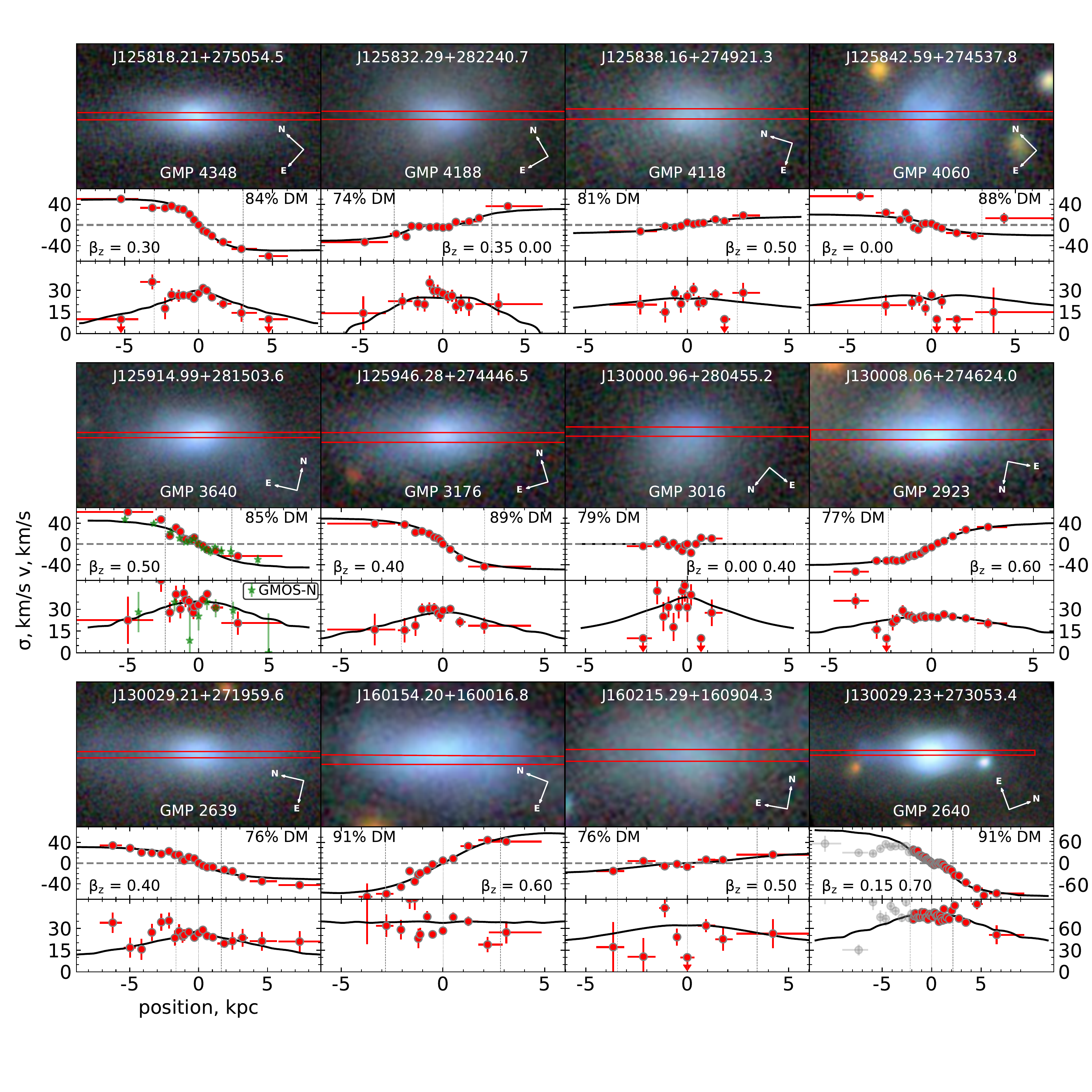}
\caption{
Top panel for each galaxy: an optical colour image with an over-plotted slit position; middle panel: stellar radial velocity $v$; bottom panel: stellar velocity dispersion $\sigma$. Black lines show the best-fitting Jeans models for stellar kinematics. The estimated dark matter contribution within $1 r_{e}$ (indicated by vertical dashed lines) and the anisotropy parameter $\beta_z$ are shown in the upper and lower corners of each $v$ panel. For 8 out of 9 galaxies $\sigma$ and $v$ profiles are shown in the same scale. For the most massive galaxy, GMP~2640 the scale of these profiles is decreased. This galaxy also has a clump of young stars whose dynamics cannot be explained in the framework of axisymmetric modelling -- the data points corresponding to this clump are greyed out.
}
\label{ps10_kin}
\end{figure*}

We simultaneously fit Binospec spectra and broadband fluxes from the far-ultraviolet ($\lambda=0.17$~$\mu$m) to the near-infrared ($\lambda$=4.5~$\mu$m) extracted from the same region against a grid of PSG stellar population models using the {\sc NBursts+phot} technique\cite{Nburstsphot} (see Extended Data Figures~ \ref{psg_spec_main}--\ref{psg_spec_sup}). The best-fitting solution yields four parameters: (i) the truncation age (i.e. end of ram pressure stripping); (ii) the fraction of gas consumed from a primordial reservoir before truncation, which is linked to the final stellar metallicity; (iii) the mass fraction of stars born in the final starburst; and (iv) a coefficient for galactic winds. In addition, we obtain radial velocity ($V_R$) and velocity dispersion of stars ($\sigma$). In the primary sample of 11 PSGs, the truncation ages range from 180~Myr to 970~Myr. In 9 of the 11 galaxies the mass fraction of stars born in the ram pressure-induced starburst lies between 10\%\ and 40\%. The fraction of consumed gas for all 11 galaxies is higher than 60\%, reaching 90--96\%\ in six cases, corresponding to mass-weighted stellar metallicities of $-1.04\dots-0.62$ dex. The estimated truncation ages for the seven galaxies with bright tails (including GMP~3016) are younger ($180\dots500$~Myr) than for the remaining five, whose tails faded close to the limit of detectability. Deriving mass-to-light ratios from our best-fitting parameters, we estimate the total stellar masses of our PSGs to be in the range 5.1$\times 10^8 M_{\odot}$ (GMP~3176) to 2.36$\times 10^9 M_{\odot}$ (GMP~2640). Therefore, despite their large sizes these are all dwarf galaxies. Truncation age profiles (see Extended Data Figure~\ref{psg_spop}) for most galaxies show weak global gradients across the disc superimposed on radially increasing gradients. The radial gradients are more pronounced for denser and more massive galaxies, e.g. GMP~4348, GMP~2640, GMP~3640. The radially increasing gradients correspond to the outside-in star formation quenching expected from the ram pressure stripping. SN feedback would cause inside-out quenching and produce radially decreasing truncation age profiles. Ram pressure strips the disc gradually from one side to the other (excepting in face-on cases), consistent with the observed truncation age gradients across the disc. The gradients across the disc are preserved for hundreds of Myr because the inner regions of all the galaxies in our sample rotate like a solid body.

Young stars boost the surface brightness and allow us to measure stellar kinematics ($V_R$ and $\sigma$) to radii as large as 2.5~$r_{e}$ (Figure~\ref{ps10_kin}), substantially further than for other low-mass galaxies beyond the Local Group. All 11 galaxies from the primary sample have $V_R/\sigma>1$ in their outer parts suggesting that they are rotationally supported systems; GMP~3016 has a nearly face-on orientation. The absence of observed UDGs with such a pronounced degree of rotational support can be explained by the lack of observations to a comparably large radius and also because the stellar rotation in the inner part of a disc slows down over time due to disc expansion. If we restrict our view of the low-mass PSGs to the $0.5\dots1.0~r_e$ typically reached for faint dEs\cite{Chilingarian09} and UDGs\cite{2019ApJ...884...79C}, only 6 have rotational velocities exceeding 10~km/s.

We fit axisymmetric Jeans anisotropic models\cite{Cappellari08} to the stellar kinematics and structural properties assuming that each galaxy resides in a spherical dark matter halo. In all cases, the modelling yielded significant dark matter fractions, from 70\%\ to 95\%\ of the total mass within $1~r_e$, and moderately anisotropic orbits, similar to those of UDGs\cite{2019ApJ...884...79C}. However, the data are of far higher quality for the diffuse PSGs. The gravitational potential in these objects is rapidly changed by the removal of a substantial fraction of the gas by ram pressure on a timescale of $30\dots70$~Myr, significantly shorter than typical dynamical times in galaxies ($200\dots350$~Myr). The rapid gravitational potential change transforms quasi-circular stellar orbits to rosette-like\cite{2017ApJ...850...99S} orbits even in a massive dark matter halo. The stellar density decreases and its radial profile becomes shallower, the orbital radial anisotropy increases, manifested by a rotational velocity drop in the galaxy central regions. The disc expansion due to slow stellar mass loss during slow passive evolution ($5\dots10$~Gyr) can be neglected when a galaxy has a massive dark matter halo. However, numerical simulations\cite{2020MNRAS.497.2786T} suggest that a dark matter halo will be gradually stripped by tidal interactions with other galaxy cluster members during flybys, which would cause the overall stellar disc expansion by about 25\%\ over 10~Gyr.

\begin{figure*}[h!]
	\centering
	\includegraphics[width=0.9\hsize]{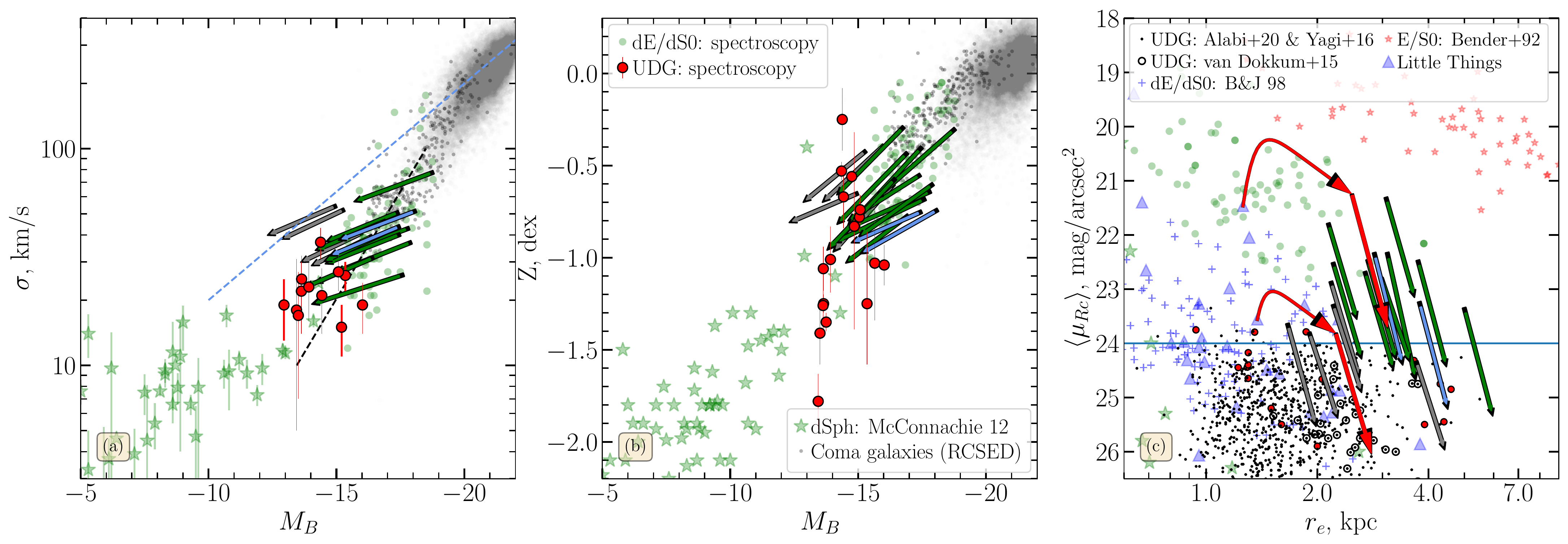}
		\caption{{\fontseries{b}\selectfont Predicted passive evolution of low-mass PSGs from their SFHs during the next 10~Gyr on the Faber--Jackson $\sigma - M_B$ (a), metallicity--luminosity $[\rm{Fe/H}] - M_B$ (b) and mean surface brightness -- effective radius $\langle \mu_{Rc} \rangle$ - $r_{e}$ (c) relations.\fontseries{m}\selectfont} We show our primary targets from the Coma and Abell~2147 clusters and additional PSGs galaxies not included in the statistical calculations by green, cyan and grey arrows respectively. Each arrow starts at the currently observed position of a galaxy and points to its expected location in 10~Gyr. On the Faber--Jackson relation we show the $L \sim \sigma^4$ and the $L \sim \sigma^{2.5}$ best-fitting linear correlations\cite{2019MNRAS.484..794G} for giant and dwarf galaxies. A cyan solid line in the right panel at $ \langle \mu_{Rc} \rangle = 24.0 \ \rm{mag}/ \rm{arcsec}^{2}$ denotes the UDG surface brightness cut. All PSGs but two will move into the UDG locus in the next 10~Gyr. The two remaining galaxies, GMP~2640 and GMP~2923 will settle in dE region. The literature sample includes dEs\cite{Chilingarian+08, Chilingarian09, 1998A&A...333...17B}, photometric\cite{2020MNRAS.496.3182A,2015ApJ...813...23V, 2016ApJS..225...11Y} and spectroscopic\cite{2018ApJ...859...37G,2018MNRAS.478.2034R,2019ApJ...884...79C, 2020MNRAS.495.2582G, 2021MNRAS.502.3144G} samples of UDGs, Local Group dwarf spheroidal galaxies\cite{McConnachie12}, Local Volume dwarf late-type galaxies from \emph{Little Things}\cite{2012AJ....144..134H}, a sample of giant early-type galaxies\cite{1992ApJ...399..462B} complemented by a selection of Coma cluster early-type galaxies (grey dots) and elliptical galaxies at $z<0.2$ (grey shaded region) from RCSED. In panel (c), the red tracks display evolutionary models of the two \emph{Little Things} objects, NGC~4214 and DDO~168 if they were ram pressure stripped: (i) a ram pressure induced starburst and disc expansion (curved tracks) followed by (ii) slow passive evolution (straight tracks).}
		\label{fjr_met}
\end{figure*}

Assuming passive evolution, based on best-fitting stellar population parameters and a slow disc expansion, we can predict internal properties (luminosity, surface brightness, luminosity- and mass-weighted stellar metallicities) of our galaxies up-to 10~Gyr from now (Figure~\ref{fjr_met}). The Faber--Jackson\cite{1976ApJ...204..668F} relation connecting stellar velocity dispersion to total luminosity and the luminosity--metallicity relation, predict that our low-mass PSGs will settle in the locus of known UDGs (7 Coma galaxies and 2 Abell~2147 galaxies) and dEs/dS0s (2 Coma galaxies) in 3--10~Gyr. The luminosity-weighted stellar metallicity will decrease over time because the contrast of the final ram pressure-induced starburst over an underlying metal-poor population will diminish. Stellar velocity dispersion will also decrease following the expansion of a galaxy. The same 9 ``future UDGs'' will also settle in the UDG locus in the $\langle \mu_{Rc} \rangle$ - $r_{e}$ diagram presenting their global structural properties\cite{Kormendy77}.

\setlength{\tabcolsep}{4pt}
\begin{table*}[h!]
\caption{\fontseries{b}\selectfont Internal dynamics, and dark matter contents of diffuse post-starburst galaxies.\label{tabgalpar}\fontseries{m}\selectfont}
{\fontfamily{cmss}\selectfont
\colorbox{tabbody}{
\begin{tabularx}{0.98\textwidth}{clcccccccccc}
No. & Name & RA & Dec & $v_r$ & $\sigma_*$ & DM$(<r_e)$ & $\beta_z$ & $\log \rho_0$ & $r_s$ & $M_{200}$ & $R_{200}$ \\
\arrayrulecolor{black}\hline
 & & deg & deg & km/s & km/s & \% & & $M_{\odot}$kpc$^{-3}$ & kpc & $10^8M_{\odot}$ & kpc \\
\arrayrulecolor{black}\hline
1 & GMP 4348 & 194.57587 & 27.84848 & 7524$\pm$1 & 51$\pm$2 & 85$\pm$8 & 0.3 & -0.5 & 1.2 & 213 & 57.8 \\
\arrayrulecolor{white}\hline
2 & GMP 3176 & 194.94282 & 27.74625 & 9690$\pm$2 & 46$\pm$4 & 88$\pm$5 & 0.4 & -0.5 & 1.0 & 123 & 48.2 \\
\arrayrulecolor{white}\hline
3 & GMP 2639 & 195.12170 & 27.33322 & 8449$\pm$1 & 43$\pm$2 & 84$\pm$3 & 0.4 & -0.8 & 1.0 & 56 & 37.2 \\
\arrayrulecolor{white}\hline
4 & GMP 4118 & 194.65901 & 27.82258 & 5350$\pm$3 & 33$\pm$7 & 80$\pm$10 & 0.5 & -1.2 & 1.5 & 66 & 39.6 \\
\arrayrulecolor{white}\hline
5 & GMP 4060 & 194.67746 & 27.76050 & 8683$\pm$1 & 26$\pm$2 & 87$\pm$15 & 0.0 & -1.2 & 2.0 & 141 & 60.4 \\
\arrayrulecolor{white}\hline
6 & GMP 2923 & 195.03358 & 27.77332 & 8665$\pm$1 & 40$\pm$1 & 69$\pm$9 & 0.6 & -1.0 & 1.5 & 112 & 47.1 \\
\arrayrulecolor{white}\hline
7 & GMP 2640 & 195.12180 & 27.51484 & 7304$\pm$1 & 78$\pm$2 & 90$\pm$4 & 0.7 & -0.4 & 2.2 & 1445 & 181.4 \\
\arrayrulecolor{white}\hline
8 & GMP 3640 & 194.81245 & 28.25101 & 7446$\pm$1 & 37$\pm$2 & 87$\pm$10 & 0.5 & -0.8 & 1.5 & 186 & 56.1 \\
\arrayrulecolor{white}\hline
9 & GMP 4188 & 194.63453 & 28.37797 & 5863$\pm$1 & 41$\pm$2 & 80$\pm$9 & 0.4 & -1.5 & 2.1 & 195 & 54.3 \\
\arrayrulecolor{white}\hline
10 & J160154.20+160016.8 & 240.47583 & 16.00467 & 11506$\pm$1 & 52$\pm$2 & 96$\pm$2 & 0.6 & -0.8 & 2.0 & 407 & 85.3 \\
\arrayrulecolor{white}\hline
11 & J160215.29+160904.3 & 240.56369 & 16.15119 & 12478$\pm$2 & 44$\pm$5 & 95$\pm$19 & 0.5 & -1.0 & 1.2 & 58 & 37.6 \\
\arrayrulecolor{black}\hline
\multicolumn{12}{l}{Additional sample}\\
\arrayrulecolor{black}\hline
12 & GMP 3016 & 195.00431 & 28.08188 & 7702$\pm$2 & 31$\pm$5 & 83$\pm$7 & 0.6 & n/a & n/a & n/a & n/a \\
\arrayrulecolor{white}\hline
13 & GMP 2673 & 195.10945 & 27.45978 & 6915$\pm$4 & 41$\pm$8 & 82$\pm$6 & 0.8 & n/a & n/a & n/a & n/a \\
\arrayrulecolor{white}\hline
14 & GMP 2552 & 195.16098 & 27.47647 & 7885$\pm$3 & 52$\pm$5 & 96$\pm$1 & 0.0 & n/a & n/a & n/a & n/a \\
\arrayrulecolor{white}\hline
15 & J160231.45+155749.9 & 240.63104 & 15.96387 & 11270$\pm$5 & 39$\pm$12 & 87$\pm$15 & 0.0 & n/a & n/a & n/a & n/a \\
\arrayrulecolor{white}\hline
16 & GMP 2662 & 195.11375 & 27.43855 & 7360$\pm$6 & 54$\pm$12 & 94$\pm$2 & 0.6 & n/a & n/a & n/a & n/a \\
\arrayrulecolor{black}\hline
\multicolumn{12}{X}{\fontsize{5.5pt}{0.9em}\fontfamily{qhv}\selectfont GMP\cite{1983MNRAS.202..113G} designations for Coma cluster galaxies and IAU-style names for Abell~2147 members (Name). Positions on the sky (RA, Dec), radial velocities ($v_r$) and stellar velocity dispersions ($\sigma_*$) obtained from the full spectrum fitting. Dynamical parameters: dark matter contribution to total mass within $r_e$ (DM),the orbital anisotropy parameter ($\beta_z$), logarithm of the central density ($\log \rho_0$) and radial scale ($r_s$) of the dark matter halo (for the main galaxy sample), dark matter halo mass ($M_{200}$) inside $R_{200}$, radius ($R_{200}$), where the dark matter density reaches the value 200 times greater than the critical density.}
\end{tabularx}
}
}
\end{table*}

\section*{Implications for evolution of ultra-diffuse galaxies}

Because our PSG sample in the Coma cluster is complete and because all galaxies were likely formed via ram pressure stripping, we can estimate the observed UDG fraction formed through this evolutionary channel. We compare the surface brightness evolution of low-mass PSGs to a complete photometric sample of Coma UDGs\cite{2016ApJS..225...11Y}. The reconstructed SFHs followed by passive evolution and slow stellar disc expansion, allow us to estimate how long each PSG will remain above the SDSS magnitude limit ($m_r<17.77$~mag) after star formation quenching, and if (or when) its surface brightness drops below the adopted UDG threshold ($\langle \mu_{e,R} \rangle > 24.0$~mag~arcsec$^{-2}$). The ratio of the duration of the UDG phase to the duration of the SDSS spectroscopic phase gives us a statistical estimate of the number of UDGs corresponding to each PSG in our sample.  We estimate that in the next 10~Gyr, 7 of the 9 low-mass Coma PSGs (not counting GMP~3016) combined correspond to 19.3$\pm$7.0 ``future'' UDGs with $\langle \mu_{e,R} \rangle > 24.0$~mag.  The remaining 2 galaxies will evolve into dEs/dS0s with higher surface brightnesses. To estimate the number of present-day Coma cluster UDGs which could have met the SDSS spectroscopic surface brightness and magnitude selection criteria any time in the past, we need to make an assumption about their SFHs.  We bound the possibilities with the extreme assumption of an instantaneous starburst 12~Gyr ago represented by a simple stellar population. This produces 90 UDGs, with 21$\pm$7\%\ formed via ram pressure stripping. If we assume a scenario suggested by our PSGs but that occurred 10~Gyr ago, constant star formation and a final starburst that formed 30\%\ of the stellar mass, we end up with 44 UDGs and a fraction of 44$\pm$16~\% formed by ram pressure stripping (see Extended Data Figure \ref{n_udg_ssp}).

\setlength{\tabcolsep}{1.5pt}
\begin{table*}
\caption{\fontseries{b}\selectfont Structural properties, Stellar population parameters and stellar masses derived from the Binospec spectra and SED modelling of diffuse post-starburst galaxies.\label{tabstpop}\fontseries{m}\selectfont}
{\fontfamily{cmss}\selectfont
\colorbox{tabbody}{
\begin{tabularx}{0.98\textwidth}{ c c c c c c c c c c c c c c}
No. & $M_{Rc}$ & $\langle \mu_{Rc} \rangle$ & $r_e$ & $t_{\mathrm{trunc}}$  & Starburst & Gas & $\lambda$ & $[Z/H]_L $ & $[\alpha/Fe]_L $ & $[Z/H]_M $ & $[\alpha/Fe]_M $ & $(M/L)_{Rc}$ & $M^*$\\
 &  & &  & & fraction & fraction &  &  &  &  &  & & \\
\arrayrulecolor{black}\hline
 & mag & mag/arcsec$^2$ & kpc & Myr & \% & \% &  & dex & dex & dex & dex & & $10^8 M_{\odot}$\\
\arrayrulecolor{black}\hline
1 & -18.21 & 22.45 & 2.73 &  825.6$\pm$3.0 &  40$\pm$2  &  98$\pm$1 & 3.7 & -0.55$\pm$0.01 & 0.09$\pm$0.02 & -0.74$\pm$0.01 & 0.09$\pm$0.01 & 0.45$\pm$0.02 & 6.0 \\
\arrayrulecolor{white}\hline
2 & -17.75 & 22.51 & 2.21 &  233.5$\pm$1.2 &  10$\pm$1  &  99$\pm$2 & 1.5 & -0.07$\pm$0.04 & 0.06$\pm$0.06 & -0.43$\pm$0.02 & 0.07$\pm$0.03 & 0.59$\pm$0.01 & 5.1 \\
\arrayrulecolor{white}\hline
3 & -18.69 & 23.34 & 4.99 &  779.3$\pm$5.5 &  60$\pm$1  &  83$\pm$1 & 1.5 & -0.45$\pm$0.01 & 0.10$\pm$0.02 & -0.58$\pm$0.01 & 0.09$\pm$0.01 & 0.39$\pm$0.01 & 7.9 \\
\arrayrulecolor{white}\hline
4 & -17.88 & 22.79 & 2.98 &  563.8$\pm$5.8 &  10$\pm$1  &  99$\pm$1 & 5.1 & -0.58$\pm$0.05 & 0.07$\pm$0.07 & -0.83$\pm$0.03 & 0.09$\pm$0.04 & 0.59$\pm$0.01 & 5.8 \\
\arrayrulecolor{white}\hline
5 & -18.43 & 22.32 & 3.09 &  413.4$\pm$2.7 &  40$\pm$2  &  47$\pm$1 & 1.5 & -0.64$\pm$0.01 & 0.05$\pm$0.01 & -0.83$\pm$0.01 & 0.06$\pm$0.01 & 0.35$\pm$0.01 & 5.8 \\
\arrayrulecolor{white}\hline
6 & -18.33 & 21.78 & 2.14 &  182.8$\pm$1.1 &  10$\pm$1  &  96$\pm$8 & 1.5 & -0.04$\pm$0.06 & 0.05$\pm$0.09 & -0.43$\pm$0.03 & 0.07$\pm$0.04 & 0.54$\pm$0.02 & 8.0 \\
\arrayrulecolor{white}\hline
7 & -19.59 & 21.31 & 3.06 &  351.3$\pm$0.5 &  20$\pm$1  &  98$\pm$2 & 2.5 & -0.28$\pm$0.03 & 0.06$\pm$0.04 & -0.58$\pm$0.02 & 0.07$\pm$0.02 & 0.50$\pm$0.02 & 23.6 \\
\arrayrulecolor{white}\hline
8 & -18.68 & 22.47 & 3.74 &  517.8$\pm$3.7 &  30$\pm$4  &  98$\pm$1 & 3.7 & -0.57$\pm$0.02 & 0.07$\pm$0.02 & -0.80$\pm$0.01 & 0.08$\pm$0.02 & 0.45$\pm$0.03 & 9.2 \\
\arrayrulecolor{white}\hline
9 & -18.06 & 22.68 & 2.69 &  400.1$\pm$4.9 &  10$\pm$1  &  55$\pm$1 & 2.5 & -0.72$\pm$0.02 & 0.06$\pm$0.02 & -0.93$\pm$0.01 & 0.09$\pm$0.02 & 0.53$\pm$0.01 & 6.1 \\
\arrayrulecolor{white}\hline
10 & -18.81 & 22.33 & 2.85 &  498.9$\pm$9.9 &  30$\pm$2  &  98$\pm$1 & 5.1 & -0.64$\pm$0.01 & 0.06$\pm$0.01 & -0.87$\pm$0.01 & 0.07$\pm$0.01 & 0.41$\pm$0.01 & 9.4 \\
\arrayrulecolor{white}\hline
11 & -18.54 & 23.23 & 3.77 &   969.9$\pm$10.9 &  60$\pm$38  &  78$\pm$3 & 1.5 & -0.43$\pm$0.12 & 0.11$\pm$0.18 & -0.56$\pm$0.04 & 0.10$\pm$0.07 & 0.41$\pm$0.21 & 7.4 \\
\arrayrulecolor{white}\hline
\multicolumn{14}{l}{Additional sample}\\
\arrayrulecolor{white}\hline
12 & -17.34 & 22.86 & 2.17 &  264.7$\pm$3.8 &  10$\pm$1  &  96$\pm$2 & 5.1 & -0.53$\pm$0.02 & 0.06$\pm$0.03 & -0.84$\pm$0.01 & 0.08$\pm$0.02 & 0.51$\pm$0.01 & 3.1 \\
\arrayrulecolor{white}\hline
13 & -17.04 & 24.33 & 3.70 &  327.2$\pm$3.1 &  5$\pm$2  &  91$\pm$1 & 1.5 & -0.25$\pm$0.02 & 0.06$\pm$0.03 & -0.51$\pm$0.01 & 0.08$\pm$0.02 & 0.69$\pm$0.01 & 3.1 \\
\arrayrulecolor{white}\hline
14 & -16.13 & 23.80 & 1.90 &  786.7$\pm$4.5 &  20$\pm$4  &  99$\pm$1 & 1.5 & -0.22$\pm$0.03 & 0.08$\pm$0.04 & -0.44$\pm$0.01 & 0.08$\pm$0.01 & 0.64$\pm$0.05 & 1.3 \\
\arrayrulecolor{white}\hline
15 & -17.25 & 23.29 & 2.20 &  245.8$\pm$3.7 &  5$\pm$1  &  99$\pm$1 & 5.1 & -0.51$\pm$0.02 & 0.06$\pm$0.03 & -0.82$\pm$0.01 & 0.09$\pm$0.01 & 0.59$\pm$0.02 & 3.2 \\
\arrayrulecolor{white}\hline
16 & -15.74 & 23.63 & 1.66 &  230.0$\pm$5.1 &  10$\pm$3  &  96$\pm$8 & 5.1 & -0.51$\pm$0.01 & 0.06$\pm$0.01 & -0.83$\pm$0.01 & 0.08$\pm$0.01 & 0.50$\pm$0.02 & 0.7 \\
\arrayrulecolor{black}\hline
\multicolumn{14}{X}{\fontsize{5.5pt}{0.9em}\fontfamily{qhv}\selectfont Galaxy number corresponding to the Table 1. Structural parameters: absolute magnitude, mean surface brightness in the $Rc$ band within the effective radius and effective radii ($\langle \mu_{Rc} \rangle, r_e$). Stellar population parameters from the full spectrum fitting: age of truncation ($t_{\mathrm{trunc}}$), fraction of stellar mass formed in the final ram-pressure-induced starburst (SB fraction), fraction of gas consumed into stars prior to ram-pressure stripping (Gas fraction), linear coefficient for galactic winds ($\lambda$). Parameters computed from the stellar population models: luminosity-weighted metallicities ($[Z/H]_L$), $\alpha$-enhancements ($[\alpha/Fe]_L$) and their mass weighted equivalents ($[Z/H]_M$, $[\alpha/Fe]_M$)}
\end{tabularx}
}
}
\end{table*}

All nine low-mass Coma PSGs have high peculiar radial velocities ($500\dots2000$~km/s) and move on non-radial orbits like many jellyfish galaxies\cite{2018MNRAS.476.4753J}. Their positions in the $\Delta v_r - d_{\mathrm{proj}}$ caustic diagram of the Coma cluster are consistent with the hypothesis that they have entered the inner part of the cluster for the first time. They have never crossed the dense cluster centre and are unlikely to ever do so, allowing them to survive for an extended period in a dense cluster environment. Observations\cite{2020ApJ...899...98V} and simulations\cite{2019MNRAS.487.5272J, 2021MNRAS.502.4262J} confirm that ram pressure efficiently strips galaxies on tangential orbits, but the longer stripping timescales increase the efficiency of star formation triggered by ram pressure-induced gas compression\cite{2020ApJ...901...95C}. During such a starburst a galaxy may reach its maximum lifetime surface brightness. GMP~4188, GMP~4118, and J160215.29+160904.3 are examples of systems on non-radial orbits with lower stellar mass and therefore low surface brightness, which barely made it into the SDSS.

For a galaxy infalling on a radial orbit, the ram pressure-induced starburst produces fewer stars due to rapid gas removal, and the stripped tail will be longer, fainter and harder to detect. Qualitatively, we can assume that for a galaxy infalling towards the cluster centre, the ram pressure becomes sufficient to ignite a starburst when it reaches some critical pressure value ($P_{SB}$); and then when it reaches $P_{S} = P_{SB}+\Delta P; \Delta P \ll P_{SB}$, the galaxy is stripped completely. The mass of stars formed should depend on the time needed to gain $\Delta P$. $\Delta P$ will be gained much faster if a galaxy approaches the cluster core on a purely radial orbit with a pericentral distance close to zero than if it approaches on an orbit with a similar eccentricity ($e\sim1$) but with a larger pericentral distance just sufficient to strip the galaxy completely. In the former case, the radial density gradient and the radial component of the orbital velocity $v_{r,\mathrm{orb}}$ work together because $P \propto \rho v_{\mathrm{orb}}^2$ and $v_{\mathrm{orb}}\approx v_{r,\mathrm{orb}}$. The large fraction of rapidly removed gas would also strongly increase the orbital anisotropy and decrease the observed stellar rotational velocity. The galaxies on radial orbits will fade and evolve into the UDG regime more quickly, but their low surface brightnesses will reduce their chances of inclusion in the SDSS spectroscopic sample anytime during their lifetime. Fainter and smaller GMP~3016, GMP~2673, GMP~2662\cite{2019ApJ...884...79C}, and J160231.45+155749.9 illustrate such a fainter continuation of the UDG evolutionary sequence. Such galaxies will also be harder to distinguish from those formed via different proposed UDG formation channels, which often produce non-rotating spheroidal stellar systems\cite{2015MNRAS.452..937Y}.

Were the late-type UDG progenitors normal-sized dwarf galaxies prior to falling onto the cluster that expanded due to ram pressure stripping? Or were they pre-processed by tidal interactions and supernovae feedback that had puffed them up first, followed by ram pressure stripping that quenched them without increasing their size? Simulations\cite{2017ApJ...850...99S} suggest that a stellar disc of a low-mass gas-rich (gas fraction 70\%) galaxy sitting in a $M_{200} = 10^{11} M_{\odot}$ dark matter halo (slightly more massive than our PSGs) would expand by a factor of $\sim$1.5 due to ram pressure stripping alone while the stellar mass loss would remain negligible. On the other hand, the ``Little Things'' survey\cite{2012AJ....144..134H} of local dwarf late-type galaxies includes 9 of 37 objects with regular kinematics having $r_e>1$~kpc, high gas fractions, and dark halo masses in the range of $10^{10.2} \dots 10^{10.7} M_{\odot}$\cite{2015AJ....149..180O}. If they fell onto a cluster, they would first become brighter because of the ram pressure induced starburst, but later, when stripped, would expand to $r_e=1.5\dots5$~kpc and fade as a result of passive evolution, finally ending up in the UDG locus of the $\langle \mu_e \rangle - r_e$ diagram as shown by red tracks for the two examples in Fig.~\ref{fjr_met}. Their largely undisturbed morphology and kinematics along with sparse environment suggest that they are ``normal'' low-mass discs, which did not undergo significant pre-processing, hence, corroborating the scenario that ram pressure stripping alone can transform many dwarf spiral and irregular galaxies into UDGs by causing both quenching and disc expansion. Because the radial distribution of neutral gas in gas-rich dwarfs is usually much shallower than stellar density profiles\cite{2012AJ....144..134H}, the ram pressure induced star formation could further grow $r_e$ by stronger stellar surface density increase in the outer parts of the disc compared to the inner regions. A search in RCSED for low-mass ($M_*<3\times10^9 M_{\odot}$) extended ($r_e>2$~kpc) star-forming galaxies within 5~Mpc from the Coma cluster centre reveals about 50 objects with the morphologies resembling those of ``Little Things'' dwarfs, which might become the progenitors of a new generation of PSGs and then UDGs, should they fall onto the Coma cluster core in the future.

Our observations of recently ram-pressure stripped low-mass disc galaxies suggest: (i) that ram pressure stripping is a viable channel of UDG formation and that UDGs formed via this channel \emph{de facto} extend the corresponding dE/dS0 sub-class to lower surface brightnesses and larger sizes; (ii) that ram pressure stripping is responsible for a substantial fraction of large rotating UDGs in galaxy clusters including extreme objects like DF~44\cite{2016ApJ...828L...6V} and also higher surface brightness, spatially extended, dwarf early-type galaxies; (iii) that ram pressure stripping affects not only morphological appearance but also stellar dynamics by increasing the stellar disc orbital anisotropy; and (iv) that the tangential orbits of UDG progenitors at large pericentral distances lead to higher ram pressure-induced star formation efficiencies, and higher stellar masses and surface densities.


\begin{methods}

\subsection{PSG selection criteria.}
We select blue extended PSGs from the RCSED\cite{RCSED} database by applying two sets of selection criteria. One of them has a stricter stellar age limit, and another one limits the integrated $g-r$ colour with slightly more relaxed stellar age restriction. This allows us to minimize the loss of possible objects of interest because of the age-metallicity degeneracy that affects the results of the full spectrum fitting of a large galaxy sample using simple stellar population models: the reported overestimated intermediate stellar age (2--3~Gyr) is compensated by the underestimated stellar metallicity, and this degeneracy can be resolved by using broad-band colours. The first set of selection criteria includes:
\begin{itemize}
\setlength\itemsep{0pt}
\item $F_{H\alpha}/\Delta F_{H\alpha}<5$ AND $F_{[O\mathrm{III}]} < 2\times 10^{-16} erg/cm^2/s$ to select galaxies without prominent emission lines.
\item Mean stellar populations younger than 1~Gyr. If a spectrum of a galaxy has low signal-to-noise ratio (SNR), its age may shift to the higher values with lower values of metallicity due to age-metallicity degeneracy. To take into account this fact we supplemented our criteria by joining with boolean {\sc OR} a metallicity condition: $\mathrm{[Z/H]} < -1.3$ dex.
\item Galaxies with spectral SNR $>$ 4 reported by SDSS, which is typically underestimated by 30--35\%. This criterion rejects spectra with unreliable signal.
\item There is no template mismatch, $\rm \chi^2_{ssp}<1$
\item We select extended objects by setting the difference between 3-arcsec fiber magnitude and integral magnitude greater than 0.5~mag in the $r$ band. This criterion selects galaxies with $r_{e} > 1.5 kpc$ at the Coma cluster distance.
\item Blue integrated colour $g-r < 0.55$~mag
\item Low redshift $0.015 < z < 0.06$
\end{itemize}

This set of criteria returned 7 objects. After visual inspection we rejected one with unreliable $H\alpha$ flux measurements and another one with unreliable integrated photometry in SDSS DR7 bringing its $g-r$ colour below 0.55~mag, when in Dark Energy Camera Legacy Survey (DECaLS\cite{2019AJ....157..168D}) it has $g-r=0.61$~mag. Finally this filter yielded 5 galaxies.

The second filter includes the following conditions:
\begin{itemize}
\setlength\itemsep{0pt}
\item $F_{H\alpha}/\Delta F_{H\alpha}<10$ AND $F_{H\alpha} < 2.5\times 10^{-16} erg/cm^2/s$ AND $F_{[O\mathrm{III}]} < 2.0\times 10^{-16} erg/cm^2/s$
\item Mean stellar populations younger than 2 Gyr.
\item $\chi^2_{ssp}<1$
\item Difference between 3-arcsec fiber magnitude and integral magnitude greater than 0.5~mag in the $r$ band.
\item Blue integrated colour $g-r < 0.4$~mag
\item Low redshift $0.015 < z < 0.06$
\end{itemize}

These criteria select 10 galaxies. We rejected one object that is apparently interacting with a companion and has incorrect photometric measurements in the SDSS/RCSED, leaving 9 galaxies. 

We joined the lists of galaxies selected using both sets of criteria. The two lists have two objects in common, leaving 12 galaxies. All belong to the SDSS main galaxy sample so that these 12 galaxies form a nearly complete magnitude limited sample within the SDSS DR7 footprint. Ten galaxies are Coma cluster members, 2 galaxies are Abell~2147 cluster members, and the remaining galaxy is a group member\cite{Saulder+16}. In this study we consider only 11 cluster galaxies, because the remaining object does not possess deep high-quality optical imaging data that can be used to assess its structural properties.

We selected additional candidate PSGs without SDSS spectroscopic confirmation as blue 5$\sigma$ outliers towards fainter near-UV magnitudes from the universal three-dimensional ultraviolet--optical colour--colour--magnitude relation for non-active galaxies\cite{CZ12} with sizes corresponding to $r_e>1.5$~kpc at the distance of their host cluster (Coma or Abell~2147). This selection returned 5 objects in Coma and 2 in Abell~2147. We placed three Coma and one Abell~2147 candidates in the Binospec slitmasks and all were spectroscopically confirmed as PSGs without current star formation and members of their host clusters.

\subsection{Spectroscopic observations and data reduction.}

We obtained spectroscopic observations of 11 PSGs using the 6.5-m MMT.  We obtained long-slit spectroscopy for each galaxy with the Binospec\cite{2019PASP..131g5004F} multi-object spectrograph and reduced the data using the Binospec data reduction pipeline\cite{2019PASP..131g5005K} optimized for extended low-surface brightness objects. Small angular distances between galaxies allow us to observe 2--3 galaxies simultaneously with the same slit mask. For nine galaxies in Coma cluster we designed 4 masks: ``Coma 1'' (GMP~2640, GMP~2639), ``Coma 3'' (GMP~4348, GMP~4118, GMP~4060), ``Coma A'' (GMP~2923, GMP~3176) and ``Coma B''  (GMP~4188, GMP~3640). ``Coma 1'' was observed in two runs: (i) December 14, 2017, during the instrument commissioning and (ii) April 10, 2018. To increase S/N both datasets were combined into one. ``Coma 3'' was observed on April 7, 2018. ``Coma A'' was observed in two consecutive nights: 4th and 5th of June 2018 with mean seeing of 1~arcsec. ``Coma B'' was observed on 8th of June 2018 with 1~arcsec seeing. Both galaxies in Abell~2147 were placed on one ``A2147'' mask which was observed on the 14th of June 2018. GMP~3016 was observed on 29th of January 2020 as a filler target for another Binospec program. In all runs we used the 1000~line~mm$^{-1}$ grating with a 1~arcsec-wide slitlets with the central wavelength of 4400--4500~\AA\ that yielded the overall wavelength coverage of 3760~\AA\ to 5300~\AA. This  configuration provides an instrumental resolution of $\sigma_{\mbox{inst}}$ $\sim$ 26$\dots$34~km/s. The total exposure time was 2h~20min for the ``Coma 1'' mask and 2h for all remaining masks. 

We also added optically selected galaxies and UDGs as filler targets to each mask. Spectra of UDGs obtained with these masks were used for dynamical modelling and dark matter contribution estimate\cite{2019ApJ...884...79C}. After the analysis of all dwarf galaxy spectra, we created an additional sample that comprised of five galaxies: 3 filler galaxies in Coma cluster and one in A2147 that met our post-starburst selection criteria. The fifth galaxy, GMP~3016, is the only galaxy from study\cite{2010AJ....140.1814Y} without detected emission inside its 27~mag isophote in the deep $R$-band Subaru imaging. There is no SDSS spectrum for this galaxy, because its total magnitude is 0.03~mag below the magnitude limit of the SDSS spectroscopic selection. These galaxies have low stellar masses and surface brightnesses and are not included in SDSS spectroscopic sample.

After the reduction of all spectroscopic data using the Binospec data reduction pipeline\cite{2019PASP..131g5005K} we obtained sky subtracted, flux calibrated, wavelength calibrated two-dimensional images with error frames for each slitlet. For the sky subtraction we used a global sky model implemented in the pipeline.

Additionally, we observed GMP~3640 using the GMOS-N spectrograph operated at the 8~m Gemini-North telescope on the 24th and 25th of June of 2017 (fast turnaround program \textit{N2017-FT-22}) with the median image quality of 0.7 and 0.9~arcsec for these two nights respectively. We used the B~1200 grating and a 0.75~arcsec slit with the wavelength setup centered at 4500~\AA\ that yields the spectral resolution  $\sigma_{\mbox{inst}}$ $\sim$ 45~km/s. The total exposure time was 3h~25min. We reduced the data using our own data reduction pipeline for GMOS data.

\subsection{Stellar population properties.}

For all PSGs we simultaneously fit optical spectra and broadband magnitudes. We take ultraviolet fluxes from the catalogs based on deep (23~ksec in near-UV and 18~ksec in far-UV) GALEX imaging data. Mid-infrared magnitudes were obtained by applying Sextractor\cite{bertin11} to Spitzer IRAC frames available from the Spitzer Heritage Archive combined into super mosaics. Optical broadband fluxes were generally taken from SDSS\cite{SDSS_DR7} and DECaLS surveys. We use deep CFHT observations in the $u^*$ band instead of shallow data in SDSS $u$. We take near-IR magnitudes from UKIDSS LAS DR10 data and CFHT $K$ and $J$ where available. We apply foreground extinction correction and K-correction\cite{CMZ10} to all magnitudes.

We use a 4-dimensional {\sc miles}-based (for spectra) and {\sc pegase.2}-based (for broad-band SEDs) grid of stellar population models\cite{PSGM_IAU355} computed for the Kroupa IMF\cite{Kroupa02} to perform stellar population analysis. We treat the chemical evolution of the stellar population  self-consistently using an analytic formulation\cite{2018ApJ...858...63C} of the ``leaky box'' model\cite{1963ApJ...137..758S, 1997nceg.book.....P, 2016JPhCS.703a2004M}. 
This model allows a fraction of heavy elements produced in stars to escape via galactic winds with the rate proportional to the current  SFR\cite{1983A&A...123..121M, 2012ceg..book.....M}. 
We model the evolution of iron and $\alpha$-elements separately, accounting for the delayed contribution from SN~Ia explosions\cite{2009A&A...501..531M}. We use spatially co-added spectra in one $r_{e}$ to maximize S/N of optical spectra . As galaxy rotation may smooth spectral details, we correct spectra for the radial velocity profile. We minimize a linear combination of $\chi^2$ values for spectrum and SED with the weight of $\chi^2_{SED}$ varying between 0.4 and 0.6 for different galaxies  against the two parameters: truncation age and fraction of consumed gas. During this procedure we model observed spectra as a template convolved with the Gaussian LOSVD and multiplied by a polynomial continuum with degree from 9 to 15. Because the spectral resolution of {\sc miles}-based models is lower than that of Binospec spectra, we convolve the data with a Gaussian to match the resolution of stellar population models. Then we scan the grid of two remaining parameters: fraction of stellar mass born in the final starburst and galactic winds coefficient $\lambda$. We obtain uncertainties for the first two parameters from the covariance matrix resulting from the minimization procedure. For the two remaining parameters we estimate the uncertainties by approximating the $\chi^2$ profile near the minimum by a quadratic function. We provide formal statistical uncertainties in Tables~\ref{tabgalpar}--\ref{tabstpop}, but for our galaxies with high SNRs, the systematic errors that originate from the choice of input spectral and SED templates are much higher than statistical ones (10--15\%\ on truncation age and $\sim$0.1~dex on metallicity and $\alpha$-enhancement).  We compare luminosity weighted metallicities computed for the SFHs with the parameters derived from our modelling with the values obtained in other studies in the 3-rd panel of Fig.~\ref{fjr_met}. For this comparison we only use spectroscopic measurements of stellar populations from the literature done using different forms of the full spectrum fitting of intermediate- and high-resolution spectra and exclude those based on SED fitting\cite{2020ApJS..247...46B} or line-strength indices\cite{2019MNRAS.484.3425M} and also obtained through excessively wide slits hence reducing the spectral resolution\cite{2018MNRAS.479.4891F}. We also excluded spectroscopic measurements of NGC1052-DF2\cite{2019ApJ...874L..12D} from our analysis despite on good quality of data and reliable measurements, because distance to this galaxy remains a matter of debate\cite{2019MNRAS.486.1192T, 2020ApJ...895L...4D}.

\subsection{Stellar kinematics.}

We use the {\sc NBursts}\cite{CPSK07} full spectrum fitting technique to measure radial velocity and velocity dispersion profiles. We use adaptive binning along the slit to reach sufficient signal-to-noise ratio to reliably measure internal kinematics (7--10 per pixel depending on the object). We use {\sc pegase.hr}-based models\cite{PSGM_IAU355} that implement PSG SFHs with self-consistent evolution of stellar metallicity.  These models are similar to those used for the stellar population analysis but do not include $\alpha$-element enhancements. The spectral resolution (R$\sim$10000) of the model exceeds Binospec's (R$\sim$4000) making it possible to measure velocity dispersion down to 10~km/s\cite{2020PASP..132f4503C}. All galaxies from the main sample exhibit some sort of regular rotation within $r_e$. GMP~4060, GMP~2640 and GMP~4188 have noticeable kinematical disturbances localised in the area of significant morphological irregularities. Detailed analysis of high-resolution HST images for the brightest galaxy in our sample, GMP~2640, revealed a large number of luminous star clusters in its central part whose asymmetric distributions on the sides of the slit causes the non-uniform slit illumination and, hence, affects the radial velocity measurements. Disturbances of kinematics profiles of GMP~4060 and GMP~4188 are also of the same origin. We notice that some simulations suggest that the ram pressure stripping by itself can lead to the displacement of the dark matter cusp in a low-mass galaxy\cite{2012MNRAS.420.1990S}, which will cause a slight lopsidedness a galaxy with a stellar mass close to the lower end of what we sample in our study. 

\subsection{Structural parameters and deep Subaru images.}

For Coma cluster galaxies we use deep $Rc$-band Subaru Suprime-Cam\cite{2002PASJ...54..833M} images (1470 -- 7980 s exposure time) to determine structural parameters. For Abell~2147 only raw archival Subaru Hyper Supreme-Cam\cite{2018PASJ...70S...1M} images are available, so we reduced and co-added them. Using GALFIT\cite{Peng10} we model each galaxy with a double S\'{e}rsic profile convolved  with the point-spread-function (PSF) obtained from Gaussian fits to non-saturated stars. We use uncertainty frames obtained from reduced, sky subtracted Subaru images for the fitting procedure, adding sky background as a constant value obtained from the sky variation analysis. We derive the deprojected effective radius for each galaxy as follows. We construct the deprojected image of the galaxy using the S\'{e}rsic components from the fit results, changing the axis ratio to 1. We then measure the radius enclosing exactly half the total flux of the deprojected image.  

We also inspected Subaru Suprime-Cam $B$-band, $V$-band, $Rc$-band, and H$\alpha$ images downloaded from the Japanese Virtual Observatory. $Rc$-band images of GMP~3176 revealed a filamentary structure 70~kpc to the west from the galaxy main body. Another galaxy, GMP~3640 has three bright filaments, emerging from its main body. Several galaxies show clumps or other irregularities in their central parts. GMP~4060 and GMP~3640 exhibit the most noticeable disturbances. However, they are substantially less significant in the $Rc$ band compared to the bluer bands ($V$ and $B$) and, therefore, they are likely caused by different stellar population properties (i.e. younger stellar age) rather than variations in the stellar mass density which would significantly affect the gravitational potential. 
For Abell~2147 we retrieved raw Hyper Supreme-Cam data in the $g$, $r$, and $i$ bands with the total exposure times of 3720, 240, and 2160~sec correspondingly from the Subaru archive and reduced them using the Hyper Suprime-Cam pipeline. In $g$-band images both Abell~2147 galaxies reveal faint tails with a complex filametary structure reaching 70~kpc in length. Both these tails point roughly towards the cluster centre. 
The brightest filaments of these tails are also detected in the $r$ and $i$ bands.

\subsection{Dynamical modelling.}  

We base our dynamical modelling algorithm on the JAM routine\cite{Cappellari08} which has been used extensively to model galactic kinematics in nearby and intermediate-redshift galaxies. It solves the Jeans\cite{Jeans22} equations assuming axial symmetry. For each galaxy we construct a suite of dynamical models that correspond to different parameter values. We use a customised version of this code, which allows us to handle a dark matter halo separately from the stars. The main steps are as follows: 
(i) We use GALFIT to fit the $R$-band Subaru image of each galaxy using a two-component S\'{e}rsic model. 
(ii) The model image of a galaxy is converted into multiple two-dimensional Gaussians (6--10 per galaxy). The same procedure is applied to a spherical Burkert\cite{burkert95} dark matter density profile, and then the two Gaussian sets are co-added to construct a model of the gravitational potential. The dark matter density profile is truncated at $R_{200}$ to prevent the divergence of the multiple Gaussian expansion procedure.
(iii) A set of Gaussians is then supplied to the modelling routine that computes the radial velocity and velocity dispersion profiles convolved with the PSF from observations for a grid of parameter values. We then scan $\chi^2$ values evaluated for each model to find the best values of following parameters: the dark matter central density $\rho_0$, halo scale radius $r_s$ and stellar vertical anisotropy $\beta_z$. We also allow the inclination $i$ to vary between $\arccos(b/a)$ and $90\deg$. The stellar mass-to-light ratio $M/L_{*}$ is fixed to the best-fitting value from the stellar population analysis. We compute grids of dynamical models with the following steps: 0.1 for $\beta_z$, 0.25 for $\log \rho_0$, 0.2~kpc for $r_s$, and $6^{\circ}$ for inclination. This spacing allows us to confidently locate the position of the $\chi^2$ minimum, but does not yield statistical uncertainties for these parameters.

We apply this method to all the galaxies in our main sample because it is more physically motivated than the traditionally used ``mass follows light'' approach. It also better reproduces the observed kinematics of our galaxies. The ``mass follows light'' approach implies that the mass and luminosity profiles of a galaxy have the same shape and differ only by a constant factor $M/L_{\mathrm{dyn}}$, which becomes an extra free parameter of the fitting procedure. We use this approach for the additional galaxy sample, because the lower quality of spectra prevents us from reliably determining the dark matter halo parameters. As we mentioned above, the kinematic disturbances in three galaxies, GMP~4060, GMP~2640 and GMP~4188 are unlikely to be of a dynamical origin. 

\subsection{Disc expansion due to ram pressure stripping.}

By selection, the galaxies in our sample already have sizes comparable to rather large UDGs being at the same time substantially more extended than typical gas-rich dwarfs in sparse environments. To estimate the importance of ram pressure stripping as a disc expansion mechanism for ``normal'' dwarf late-type galaxies infalling on-to a massive galaxy cluster like Coma, we took two galaxies from {\sc Little Things}\cite{2012AJ....144..134H} sample: NGC~4214 having stellar mass comparable to GMP~2640, the most massive galaxy in our sample, and DDO~168 comparable to the smallest galaxy in the additional sample. Because the ram pressure timescale is shorter than the rotation period at $r_e$ by a factor of a few, we estimate the average disc expansion using a formula for the change of the orbital radius in case of the rapid mass loss\cite{1980ApJ...235..986H} $r_f / r_i = \frac{\varepsilon}{2\varepsilon -1}$. In this formula, $\varepsilon = M_f/M_i$, $M_i$ is initial total mass inside $r_e$ that includes dark matter\cite{2020ApJS..247...31L}, stars and gas; $M_f$ is the same without gas. This approach suggests that there is no change in the outer surface brightness profile slope, but in the inner parts the dark matter contribution decreases causing the orbital expansion to a higher degree than in the galaxy's outskirts. This creates a large flat core in the surface brightness profile and also increases $r_e$. 

Our modelling shows that ram pressure stripping is capable in expanding a stellar disc by a factor of 2 and even more and might be solely responsible for transforming compact gas-rich galaxies to quiescent stellar systems with UDG sizes and densities. We model the ram pressure induced starburst with a constant SFH for 60~Myr so that the stellar mass formed in it is 20\% of the total final stellar mass, the rest of the gas is stripped. We expect that the disc expansion initiated by ram pressure stripping will continue for 300~Myr, which roughly corresponds to the rotation periods of the galactic discs in our PSGs as well as in NGC~4214 and DDO~168. Both, the starburst and the disc expansion start simultaneously but the expansion lasts longer -- these stages correspond to the curved parts of the red tracks with an arrow at the end in Fig.~\ref{fjr_met}. Then the galaxies are left to evolve passively for 10~Gyr with a slow disc expansion due to stellar evolution (straight portions of the tracks in Fig.~\ref{fjr_met}.) At the end of the first stage of the disc expansion, 300~Myr after they entered the cluster DDO~168 and NGC~2414 would resemble GMP~3016 and GMP~2640 respectively.

\subsection{Slow disc expansion modelling.}

After a stripping event, the stars evolve, and the most massive of them will die. We assume that the galaxy will stay in the cluster after stripping and the final starburst, and that the gas released as a result of stellar evolution will be gradually stripped and will not form new stars. Therefore, the stellar mass decrease with time, leading to disc expansion.

We estimate the expansion of the disc assuming angular momentum conservation, which gives us the following expression:  
\begin{equation}
v(M_{tot}(t_0),r(t_0))\cdot r(t_0) = v(M_{tot}(t_0 + \delta t),r(t_0 + \delta t))\cdot r(t_0 + \delta t).  
\end{equation}
For $\delta t \xrightarrow{} 0$ we get the exact expression for $\dot{r}$: 
\begin{equation}
\dot{r}/r = \frac{\partial{M_{tot}}}{\partial{t}}\frac{\partial{v}}{\partial{M_{tot}}} / \left(\frac{\partial{v}}{\partial{r}}r + v\right) 
\end{equation}
This formula can be expressed in terms of logarithmic derivatives: 
\begin{equation}
\dot{r}/r = \alpha (r) \cdot \dot{M}/M;\;
\alpha (r) = M_{tot} \frac{\partial{v}}{\partial{M_{tot}}} / \left (\frac{\partial{v}}{\partial{r}}r + v \right )
\end{equation}
Applying it to an exponential disc, we find that $\alpha (r)$ is a smooth function with values between 1.3 and 3.0, reaching the maximum (3.0) at $r=0$ and $r \xrightarrow{} \infty$, with the minimum at $2 r_{e}$. The central disc expansion is therefore larger than in the outskirts. Therefore, the slope of light profile will decrease with time and the Sersic index will drop to values seen in ordinary ultra-diffuse galaxies.

The presence of significant amounts of dark matter keeps the relative change of $M_{tot}$ and its derivative ${\partial{M_{tot}}}/{\partial{t}}$ quite small, driving the full expression for $\dot{r}/r$ to near-zero values. This leads to a decrease of overall disc expansion from 20\% with no dark matter case to 3--5\% for PSGs with dark matter fractions of 70--90\%. This small value cannot affect mean surface brightness by more than 0.1~mag~arcsec$^{-2}$. However, numerical simulations suggest that dark matter haloes of low-mass galaxies can be efficiently stripped by tidal forces during close passages near other cluster members, therefore one would still expect a slow disc expansion over time reaching the same 20--25\%\ over 10~Gyr. Strong tidal interactions, which substantially disturb the galaxy morphology are rare so they generally have little effect on the UDG shapes or leave visible tidal features\cite{2017ApJ...851...27M}. The disc expansion will lead to the decrease of the average stellar surface density in the disc by 40--50\%\ without substantial disc thickening. In addition, the stellar volume density also decreases because of the slow stellar mass loss reaching 20\% over 10~Gyr. These two factors combined will lower the stellar velocity dispersion by $\sim$35\%\ because $\sigma$ is proportional to the square root of the local stellar density.

Simulations\cite{2020MNRAS.497.2786T} also show that disc expansion causes the dark matter density decrease by some 40\% in the central parts of galaxies what leads to the drop of the circular velocity by 20\%, because it is proportional to the square root of combined density of stars and dark matter, which dominates the mass. Combined with the stellar density decrease, this effect reaches $\sim$25\%. However, observed stellar rotational velocity is smaller than the circular velocity at a given position in a galaxy due to asymmetric drift ($v_a$) and to estimate the change of its value we applied the formula 4.228 from Binney \& Tremaine\cite{2008gady.book.....B} under the assumption that during the process of slow expansion the ratios of the different velocity dispersion components do not change. In this case, the change of the asymmetric drift velocity is proportional to $\sigma_R^2/v_c$. For 85\% of dark matter inside $1~r_e$ the decrease of stellar rotational velocity will be 25\%.

\subsection{UDG number statistics.}
We assess the importance of the ram pressure stripping scenario to other UDG formation channels.  We statistically estimate how many present-day UDGs correspond to each PSG in our primary sample of 9 Coma cluster galaxies and compare the result to the observed number of UDGs in the Coma cluster. We model the surface brightness evolution from the reconstructed SFH and calculate the two epochs: (i) $t_{\mathrm{SDSS}}$ when a galaxy becomes too faint to be included in the SDSS spectroscopic sample ($r_{\mathrm{tot}}>17.77$~mag) and (ii) $t_{\mathrm{UDG}}$ when it dims enough to be classified as a UDG. UDG classification has a surface brightness cutoff of $\langle \mu_R \rangle_{\mathrm{eff}}>24$~mag/arcsec$^2$. Statistically speaking, a given PSG will correspond to the number of UDG estimated as $N_{\mathrm{UDG}}=(t_{\mathrm{SDSS}}-100 \mathrm{Myr})/(10 \mathrm{Gyr} - t_{\mathrm{UDG}})$ if $t_{\mathrm{UDG}}<10$~Gyr. At 100~Myr emission lines disappear from a PSG spectrum, and 10~Gyr is the reference passive evolution time. At the end we co-add $N_{\mathrm{UDG}}$ for all 9 galaxies in the primary sample. We find that only 7 galaxies reach the UDG regime within 10~Gyr. The total number of expected UDGs from these 7 progenitors is 19.34$\pm$7.00. This estimate is made assuming a constant rate of infall of galaxies on a cluster. Simulations\cite{2013MNRAS.432.1162L} show that during the first 2--3 Gyr of the cluster formation epoch, the infall rate might be up-to twice as high as now. This may only increase our estimate of UDGs formed via the ram pressure stripping channel. The uncertainty is estimated by assuming the Poisson process of the galaxy infall and propagating stellar population parameter statistical errors and the estimated 10\%\ systematic errors for the effective radii\cite{2020MNRAS.496.3182A} and truncation ages\cite{2018ApJ...858...63C}.

We now need to estimate the total number of present-day UDGs in a complete magnitude limited sample\cite{Koda15,2016ApJS..225...11Y} with sufficient stellar mass so that 100~Myr after star formation quenching they would be bright enough to meet the SDSS spectroscopic selection. Our initial PSG selection criteria includes a minimum 0.5~mag difference between the 3-arcsec fiber magnitude and the total magnitude.  This corresponds to a lower $r_e$ limit of $\approx$1.5~kpc for galaxies at the distance of the Coma cluster.  During passive evolution,  these galaxies will expand by $\approx$25\%, so we only count the present-day UDGs having $r_e>1.875$~kpc. To estimate the stellar mass we need to assume an SFH in a present-day UDG, which we cannot infer directly from the available data. The most extreme case would be an instantaneous starburst modelled by a simple stellar population. Such a SFH leads to the minimal possible stellar mass cutoff and, hence, the largest number of UDGs passing the selection. For a 10~Gyr age, there are 90 galaxies in UDG sample\cite{2016ApJS..225...11Y} which could satisfy the selection criteria for our PSGs in the past at some moment during their lifetime.  Our lower limit of of UDGs is therefore 21$\pm$7~\%, which could be explained by the ram pressure stripping formation scenario.

For a more realistic estimate we consider a typical observed SFH for our PSGs:  a truncated constant SFR started 13.0~Gyr with 30\%\ of the total stellar mass formed in the final starburst 10~Gyr ago. This corresponds to the median mass fraction we found in the PSG analysis. The number of UDGs that could possibly satisfy our selection criteria decreases to 44, increasing the importance of the ram-pressure-stripping UDG formation channel to 44$\pm$17~\%. This fraction would increase even more if we take UDG SFHs with a lower final starburst fraction. In Extended Data Figure~\ref{n_udg_ssp}  we show how the adopted SFH affects the total number of UDGs from the study\cite{2016ApJS..225...11Y} that satisfy the SDSS spectroscopic selection criteria and our selection criteria anytime during their lifetime. We consider three SFHs, all starting 13 billion years ago  with truncation ages of 10, 8 and 5 billion years. In all cases we estimate the number of UDGs as described above. In the case of earlier truncation time, M/L ratio and stellar mass change significantly. Given the SDSS magnitude limit we have a individual magnitude limit of UDG for each scenario of star formation. If a galaxy was bright enough 10 billion years ago  to pass the selection criteria, this galaxy can now be very faint, and many galaxies from the sample of present-day UDGs are brighter than this limit. In case of more extended scenario of star formation, the M/L ratio and stellar mass change less, we infer a significant fraction of long-lived old stars. The magnitude limit for UDGs is correspondingly brighter, decreasing the number of UDGs. The most extreme case is a SFH with truncation age of 5 Gyr without a final starburst, i.e. constant SFR from 13~Gyr to 5~Gyr. In this case none of known UDGs pass the SDSS selection criteria.

\end{methods}

\begin{datacodeavailability}
Reduced flux calibrated Binospec spectra co-added within 1$r_e$ and thier best-fitting {\sc NBursts+phot} spectrophotometric models (see Extended Data Figures~\ref{psg_spec_main}--\ref{psg_spec_sup}), the spatially resolved profiles of internal kinematics and stellar populations and corresponding axisymmetric Jeans models for every galaxy as are available through {\sc Zenodo} data repository with the following permanent identifier: \href{https://doi.org/10.5281/zenodo.5031351}{\textcolor{natastteal}{https://doi.org/10.5281/zenodo.5031351}}. That package also contains a compilation of data points used in Fig.~\ref{fjr_met} and a {\sc python} procedure to reproduce the figure. Fully reduced Subaru and CFHT images used for the dynamical modelling procedure are available in the corresponding open access data archives (\href{http://jvo.nao.ac.jp/index-e.html}{\textcolor{natastteal}{http://jvo.nao.ac.jp/index-e.html}}; \href{http://www.cadc-ccda.hia-iha.nrc-cnrc.gc.ca/en/}{\textcolor{natastteal}{http://www.cadc-ccda.hia-iha.nrc-cnrc.gc.ca/en/}}).
\end{datacodeavailability}

\vskip 12pt
\noindent
Received: 24 October 2020; Accepted: 27 July 2021;\\ 
Published online: 01 November 2021

\bibliography{PSG}

\begin{addendum}
 \item ~\\ KG, IC, AA, IK acknowledge the Russian Science Foundation (RScF) grants No. 19-12-00281 for supporting the search of UDG progenitors and data analysis of a multi-wavelength spectrophotometric dataset, and No. 17-72-20119 for supporting the development of stellar population models used in the data analysis; and the Interdisciplinary Scientific and Educational School of Moscow University ``Fundamental and Applied Space Research''. KG acknowledges the support from the Foundation of development of theoretical physics and mathematics ``Basis''. IC's research is supported by the Telescope Data Center at Smithsonian Astrophysical Observatory. We thank D.~Eisenstein, F.~Combes, G.~Mamon, and O.~Sil'chenko for fruitful discussions related to this project. We are grateful to the staff of the MMT Observatory jointly operated by Smithsonian Astrophysical Observatory and the University of Arizona for their support of Binospec operations and service mode observations. This study is based in part on data collected at Subaru Telescope and retrieved from the SuprimeCam and Hyper-SuprimeCam data archive system, which is operated by Subaru Telescope and Astronomy Data Center at National Astronomical Observatory of Japan and on observations obtained at the international Gemini Observatory (proposal GN-2017A-FT-22), a program of NSF’s NOIRLab, which is managed by the Association of Universities for Research in Astronomy (AURA) under a cooperative agreement with the National Science Foundation on behalf of the Gemini Observatory partnership: the National Science Foundation (United States), National Research Council (Canada), Agencia Nacional de Investigaci\'on y Desarrollo (Chile), Ministerio de Ciencia, Tecnolog\'ia e Innovaci\'on (Argentina), Minist\'erio da Ci\^encia, Tecnologia, Inova\c{c}\~oes e Comunica\c{c}\~oes (Brazil), and Korea Astronomy and Space Science Institute (Republic of Korea).
 \item[Author Contributions] ~\\ K.G. computed stellar population models, prepared Binospec observations, analyzed Binospec data and wrote the manuscript; I.C. formulated PSG search criteria and discovered the galaxies in the RCSED catalog, reduced Binospec data, modified the full spectrum fitting code to analyze them and wrote the manuscript; A.A. performed surface photometry, performed dynamical modelling and wrote the manuscript; D.F. led the development of the Binospec spectrograph and proposed the observational strategy; I.K. jointly with I.C. developed the spectrophotometric fitting code {\sc NBursts+phot} and validated the spectroscopic data analysis; S.M. supported preparation and execution of observations; M.Y. provided cutouts of deep Subaru images and did the reduction of archival Subaru data for Abell~2147; I.C., D.F., and S.M. performed Binospec observations during the commissioning and science verification of the instrument; all authors discussed the manuscript.
 \item[Competing Interests] ~\\ The authors declare no competing interests.
 \item[Additional information] ~
    \\ {\bf Extended data} is available for this paper at \href{https://doi.org/10.1038/s41550-021-01470-5}{\textcolor{natastteal}{https://doi.org/10.1038/s41550-021-01470-5}}.
    \vspace{4pt} \\ {\bf Correspondence and requests for materials} should be addressed to Igor Chilingarian (email: igor.chilingarian@cfa.harvard.edu; chil@sai.msu.ru) and Kirill Grishin (email: grishin@voxastro.org).
    \vspace{4pt} \\ {\bf Peer review information Nature} Astronomy thanks Jeffrey Kenney and the other, anonymous, reviewer(s) for their contribution to the peer review of this work.
    \vspace{4pt} \\ {\bf Reprints and permissions} information is available at \href{www.nature.com/reprints}{\textcolor{natastteal}{www.nature.com/reprints}}.
    \vspace{4pt} \\ {\bf Publisher's note} Springer Nature remains neutral with regard to jurisdictional claims in published maps and institutional affiliations.
\end{addendum}

\clearpage

\begin{Extended Data Figure*}
\centering
\includegraphics[width=1.00\hsize]{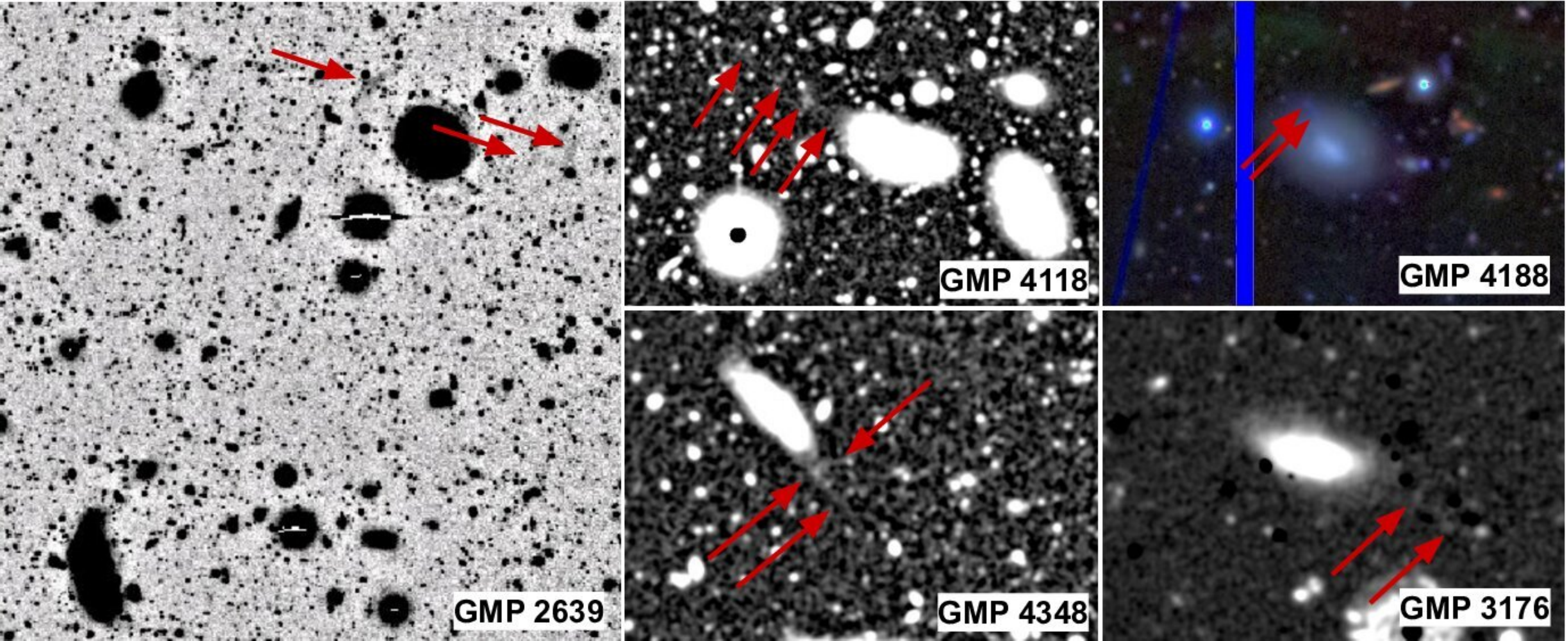}
\caption{Low-contrast filaments suspected to be ram-pressure-stripped tails for five Coma galaxies. For GMP~2639 we show an inverted $Rc$-band image binned with a 4$\times$4~pix window. For GMP~4118 and GMP~4348 we show $H_{\alpha}$ images without continuum subtraction convolved with a 2-D Gaussian kernel having FWHM in range $5\dots7$~pix depending on S/N of input images. For GMP~3176 we show a difference between $B$ and $Rc$ Subaru frames. A spot at the end of the short filament in GMP~4188 pointed by an arrow is a bright H$\alpha$ source.}
\label{psg_suspected_tails}
\end{Extended Data Figure*}

\begin{Extended Data Figure*}
\centering
\includegraphics[clip,trim={0.0cm 0.9cm 0.4cm 0.5cm},width=0.85\hsize]{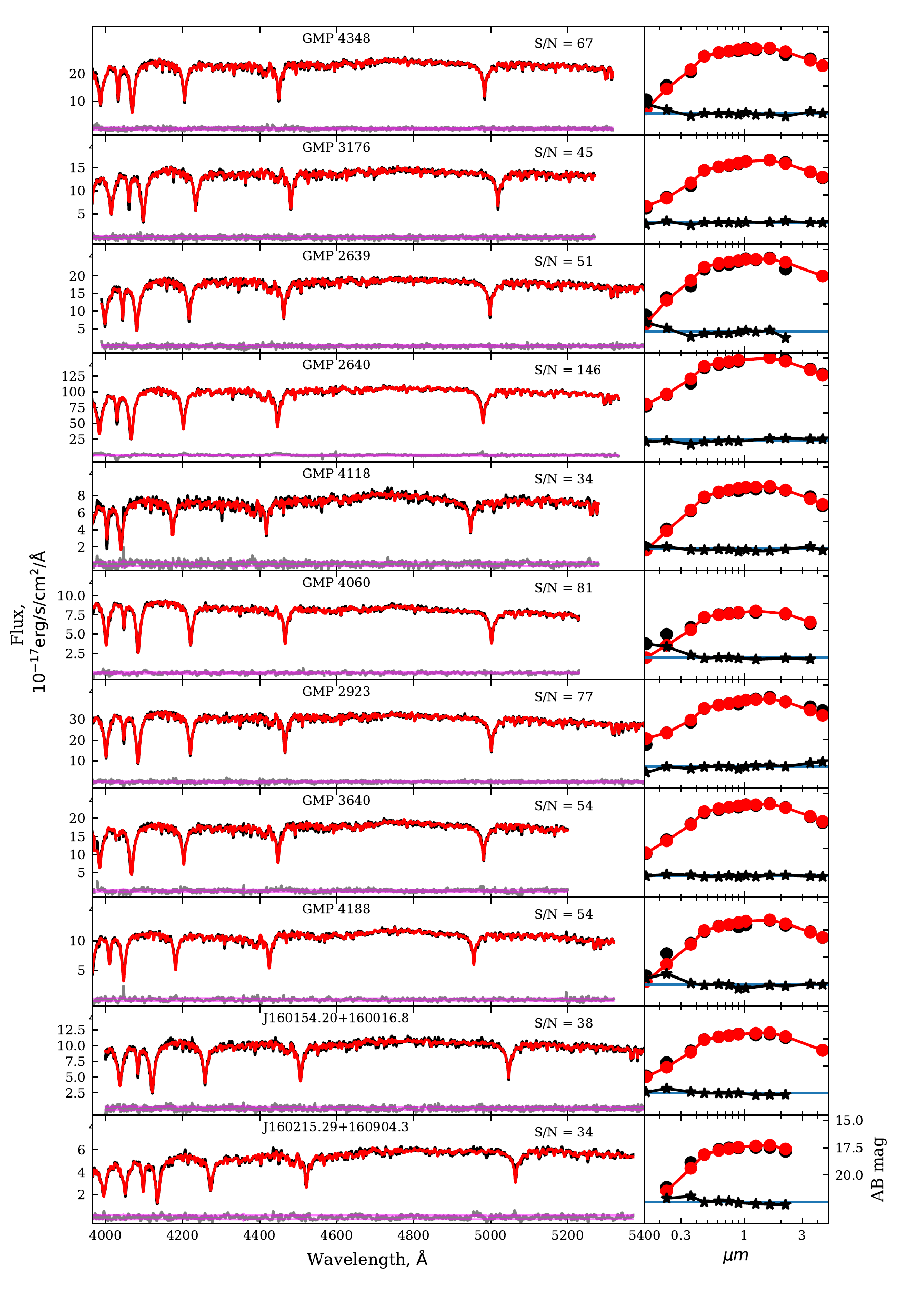}
\caption{Observed data (black) degraded down to correspond template resolution and best-fitting templates (red) for Binospec spectra (left panels) and broad-band SEDs (right panels) for PSGs from the main sample. For the spectra we also show $1 \sigma$ uncertainty profiles (magenta) and fitting residuals (grey). SED panels show observations and best-fitting models as black and red dots correspondingly. The residuals are shown with black stars around a cyan line representing zero level in the same scale as SEDs. The upper right corner of each spectrum contains a median signal-to-noise ratio for a degraded spectrum per pixel. Main spectral features and broad-band filter bandpasses are marked in the upper row of panels.}
\label{psg_spec_main}
\end{Extended Data Figure*}

\begin{Extended Data Figure*}
\centering
\includegraphics[clip,trim={0.4cm 0.0cm 0.4cm 0.2cm},width=0.85\hsize]{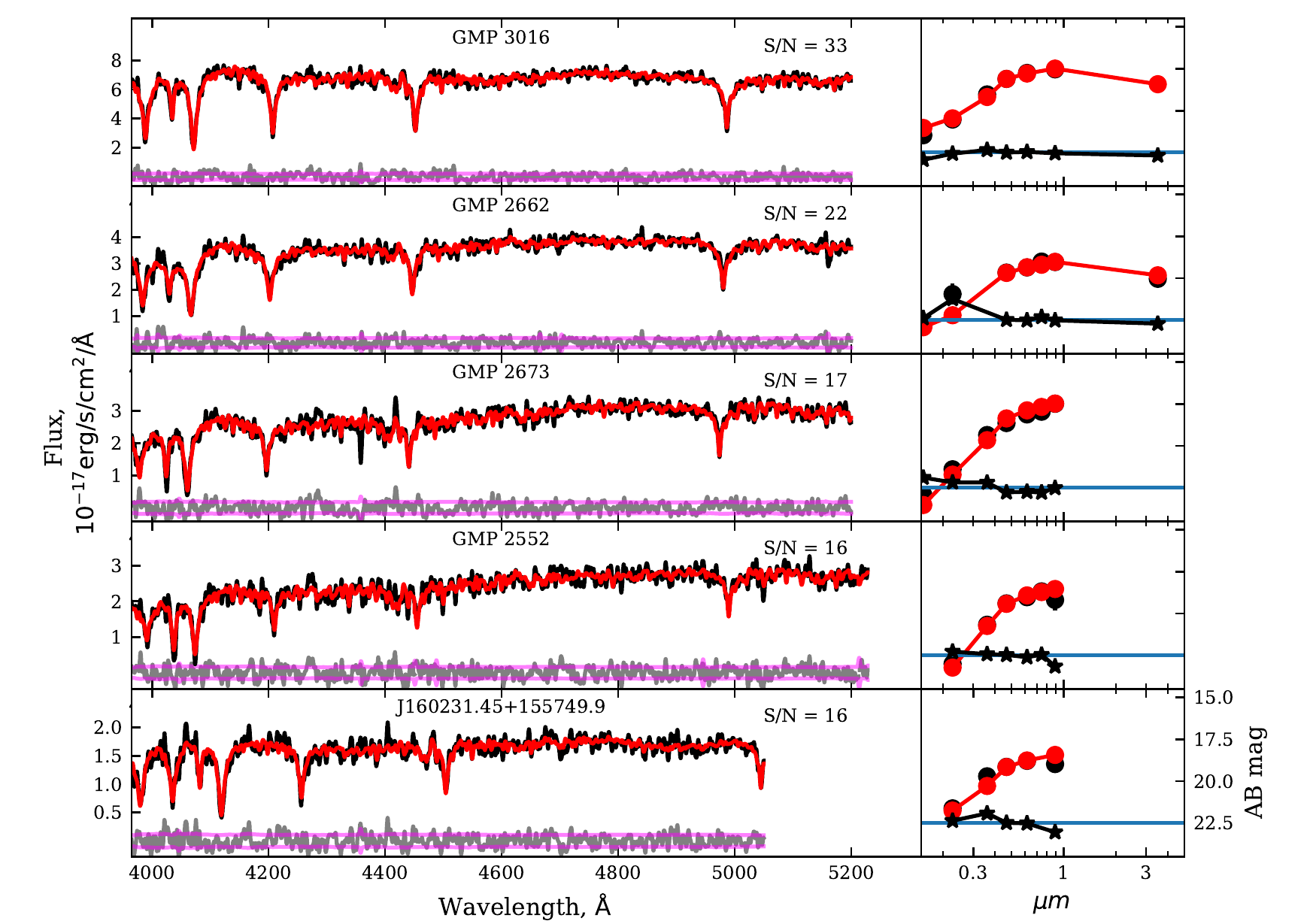}
\caption{The same as Extended Data Figure \ref{psg_spec_main} for the additional galaxy sample.}\label{psg_spec_sup}
\end{Extended Data Figure*}

\begin{Extended Data Figure*}
\centering
\includegraphics[width=\hsize]{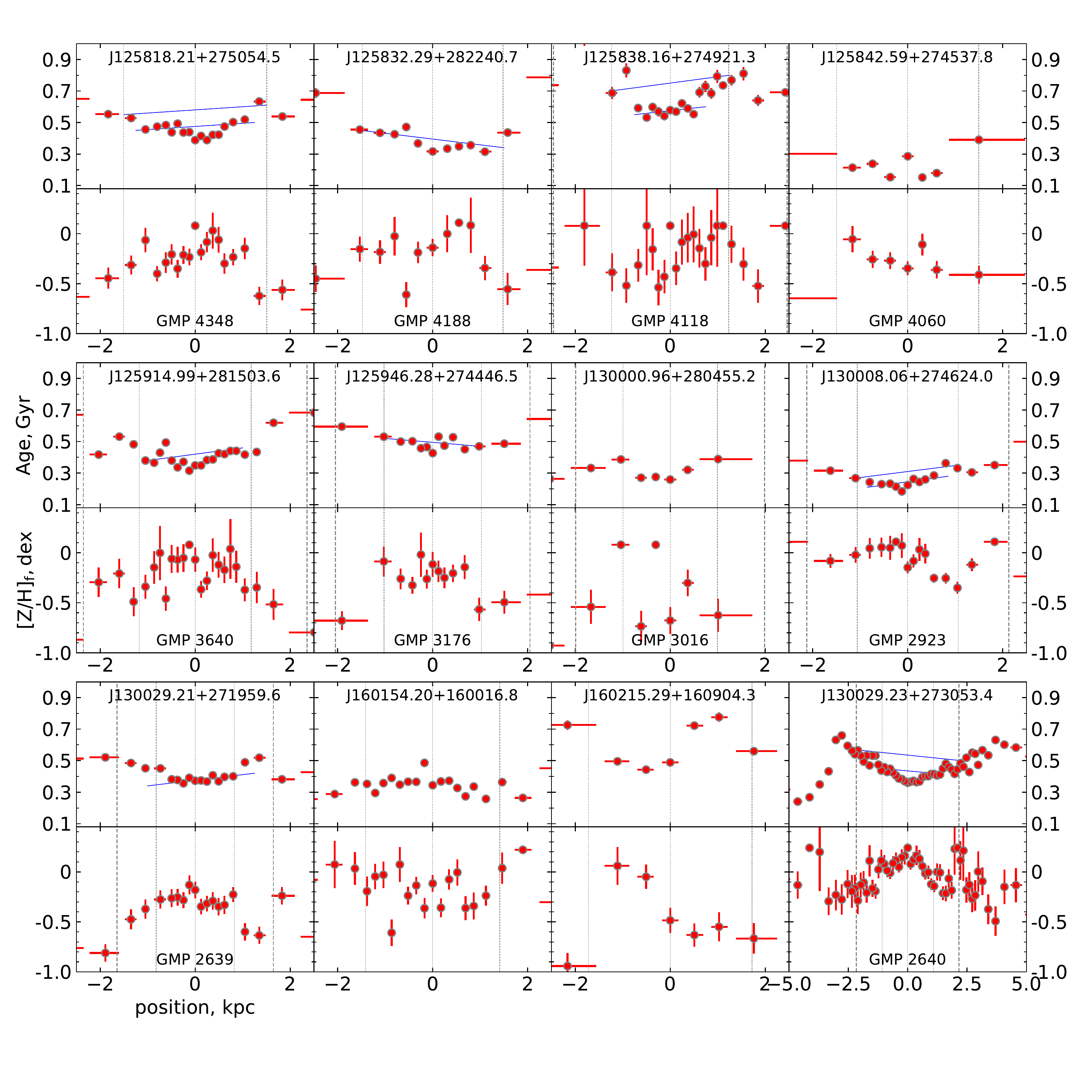}
\caption{Resolved stellar population profiles for the PSGs from the main sample and GMP~3016 obtained from full spectral fitting of long-slit spectra using {\sc pegase.hr}-based templates\cite{PSGM_IAU355}. We perform the $\chi^2$ minimization varying the truncation age and final metallicity which describes the metallicity of final starburst while other parameters (wind coefficients and SB fraction) are fixed to values derived from the modelling of the co-added 1D spectra with broadband photometry. For galaxies with non-radial age gradients likely caused by the ram pressure stripping, solid blue lines show the gradient trend sometimes over-imposed on the radial gradient illustrating the outside-in quenching. Dotted and dashed lines mark $\pm r_e/2$ and $\pm r_e$ respectively.}
	\label{psg_spop}
\end{Extended Data Figure*}

\begin{Extended Data Figure*}[h!]
\vskip -1cm
\centering
\includegraphics[width=0.8\hsize]{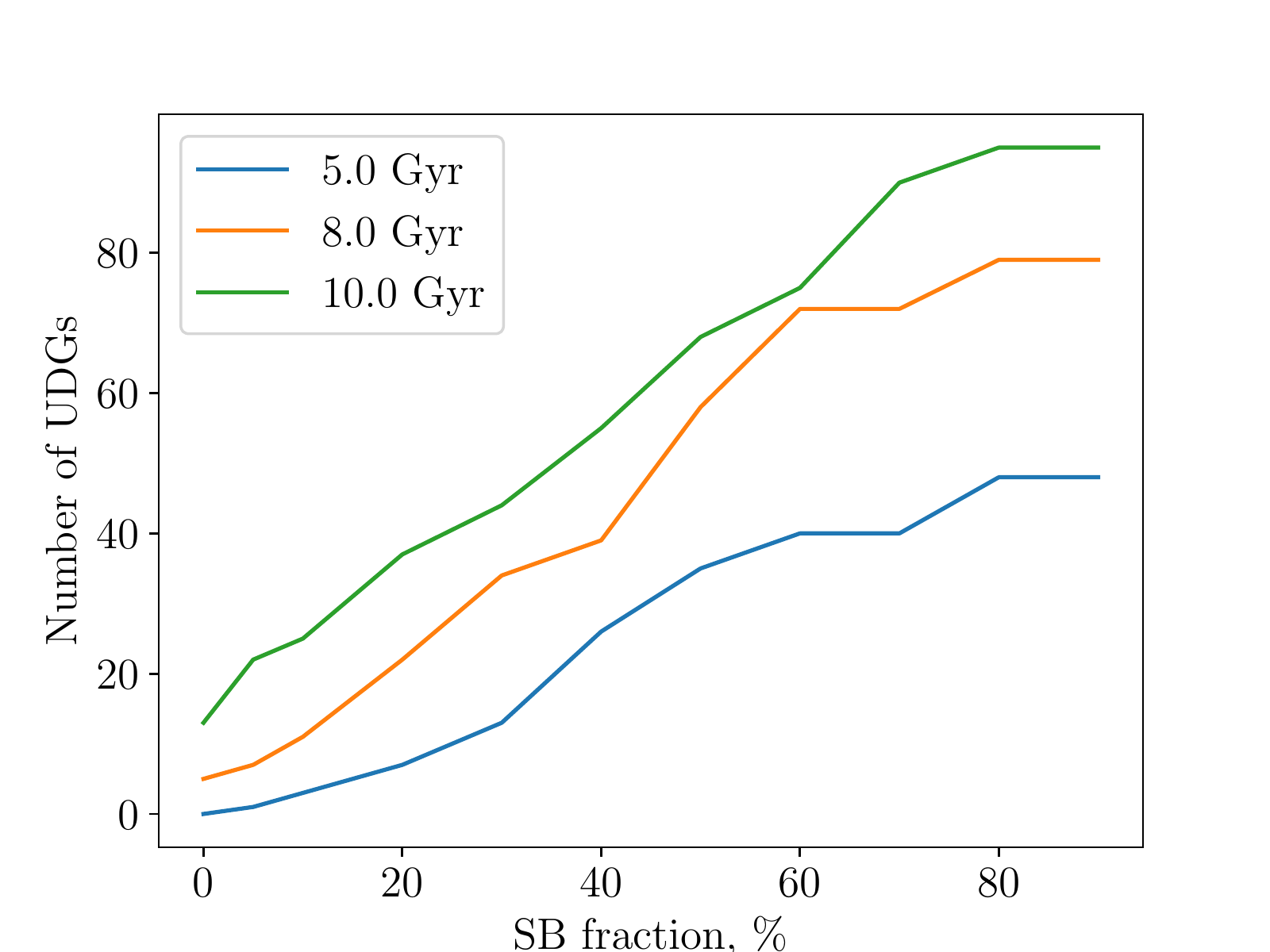}
\caption{A number of UDGs from the Subaru Coma cluster survey\cite{2016ApJS..225...11Y} that could pass the SDSS spectroscopic selection criteria and satisfy our PSG selection criteria at some moment during their lifetime as a function of a mass fraction of stars born in the final star burst for three different truncation ages: 5, 8, and 10 Gyr.}
\label{n_udg_ssp}
\end{Extended Data Figure*}

\end{document}